\documentclass[fleqn,usenatbib]{mnras}

\usepackage{newtxtext,newtxmath}

\usepackage[T1]{fontenc}

\DeclareRobustCommand{\VAN}[3]{#2}
\let\VANthebibliography\thebibliography
\def\thebibliography{\DeclareRobustCommand{\VAN}[3]{##3}\VANthebibliography}

\usepackage{graphicx}	
\usepackage{amsmath}	
\usepackage{amssymb}
\usepackage{subcaption}
\usepackage{lineno}
\usepackage{graphicx}
\usepackage{mathrsfs}

\usepackage{orcidlink}

\usepackage{pifont}

\graphicspath{{Figures/}}

\newcommand{\standard}{\textit{standard}}
\newcommand{\CRs}{\textit{+CRs}}

\newcommand{\Mtwohundred}{M_{200\mathrm{c}}}
\newcommand{\Rtwohundred}{R_{200\mathrm{c}}}
\newcommand{\Msolar}{\mathrm{M}_\odot}

\newcommand{\Xcr}{X_\mathrm{CR}}
\newcommand{\noalfven}{\textit{+CRs -- Alfv\'{e}n}}
\newcommand{\bcdot}{\boldsymbol{\cdot}}
\newcommand{\bnabla}{\boldsymbol{\nabla}}

\newcommand{\cmark}{\ding{51}} 
\newcommand{\xmark}{\ding{55}} 

\title[Impact of CRs on Disc Sizes and Outflows]{Unveiling the Impact of Cosmic Rays on the Disc Sizes and Outflows from Dwarf Scales to Galaxy Groups}
\author[R. Bieri et al.]{Rebekka Bieri$^1$\thanks{Corresponding author: Rebekka Bieri (rebekka.coles-bieri@uzh.ch )}\orcidlink{0000-0002-4554-4488},
R\"udiger Pakmor$^2$\orcidlink{0000-0003-3308-2420},
Freeke van~de~Voort$^3$\orcidlink{0000-0002-6301-638X},
Rosie Y. Talbot$^2$\orcidlink{0000-0001-9393-7879},
\newauthor
Maria Werhahn$^2$\orcidlink{0000-0003-4984-4389},
Christoph Pfrommer$^4$\orcidlink{0000-0002-7275-3998}, 
Volker Springel$^2$\orcidlink{0000-0001-5976-4599}
\vspace*{0.1cm}\\
$^{1}$Institut f\"ur Astrophysik, Universit\"at Z\"urich, Winterthurerstrasse 190, 8057 Z\"urich, Switzerland\\
$^{2}$Max-Planck-Institut f\"{u}r Astrophysik, Karl-Schwarzschild-Str. 1, 85748 Garching, Germany\\
$^{3}$Cardiff Hub for for Astrophysics Research and Technology, School of Physics and Astronomy, Cardiff University, Cardiff CF24 3AA, UK\\
$^{4}$Leibniz-Institut f\"{u}r Astrophysik Potsdam (AIP), An der Sternwarte 16, 14482 Potsdam, Germany
}

\date{Accepted XXX. Received YYY; in original form ZZZ}

\pubyear{}

\begin{document}
\label{firstpage}
\pagerange{\pageref{firstpage}--\pageref{lastpage}}
\maketitle

\begin{abstract}
Cosmic rays (CRs) are a non-thermal energy component in the interstellar and circumgalactic medium (CGM) that can act as an additional feedback channel beyond thermal and kinetic feedback from stars and AGN. They influence galaxy evolution by altering gas properties, regulating star formation, and shaping galactic outflows. We investigate these effects using a suite of cosmological zoom-in simulations, which incorporate CR transport and feedback on top of the Auriga galaxy formation model. Our simulations span a wide range of halo masses, from dwarf galaxies to small groups ($\Mtwohundred= 10^{10} - 10^{13}~\Msolar$), allowing us to assess the mass-dependent impact of CRs in a cosmological setting. We find that CRs have the strongest impact in lower-mass galaxies ($\Mtwohundred < 10^{12}~\Msolar$), where they suppress star formation rates by up to 50\%, reduce gas and stellar half-light radii, and drive outflows that reach higher velocities at the virial radius compared to simulations without CRs. These CR-enhanced outflows also transport more metals and magnetic fields into the CGM, leading to higher CGM metallicities, stronger magnetisation, and lower CGM temperatures. In more massive galaxies, CRs do not significantly affect star formation or outflow properties, likely because thermal and kinetic feedback from stars and AGN dominate in this regime. However, CRs still influence galaxy morphology across all halo masses by reducing the gas half-mass radius and stellar half-light radius. Finally, variations in CR transport properties, such as different diffusion coefficients or the exclusion of Alfvén cooling, significantly affect star formation, CGM properties, and outflows in lower-mass galaxies. This sensitivity makes these galaxies key environments for testing CR transport models and refining our understanding of their role in galaxy evolution.
\end{abstract}




\section{Introduction}

Galaxy formation remains a complex problem in astrophysics, involving a wide range of scales and physical processes. Accurately modeling galaxy evolution requires a comprehensive understanding of various factors that influence the lifecycle of a galaxy. A key element in this process is feedback from stellar winds, radiation, supernovae, and active galactic nuclei (AGN), which plays a crucial role in shaping galaxies. These feedback processes regulate star formation, drive galactic winds, and shape the properties of the galaxies. Numerical simulations provide a powerful tool to investigate how different feedback mechanisms interact and influence galaxy evolution over cosmic time. Cosmological simulations, in particular, enable us to study the formation and evolution of galaxies within their large-scale environment. While these simulations reproduce many observed galaxy properties, the physical mechanisms driving galactic winds and outflows remain an area of active research \citep{Naab+Ostriker2017, Crain+VanDeVoort2023, Feldmann+Bieri2025}. 

For low-mass galaxies, the primary driver of galactic winds is a combination of stellar radiation, stellar winds, and supernovae, \citep{Dekel+Silk1986, Efstathiou2000}. In contrast, for high-mass galaxies, outflows are thought to be predominantly driven by AGNs associated with supermassive black holes \citep[e.g.][]{Silk+Rees1998, Benson+2003}. Both stellar and AGN feedback processes generate thermal and mechanical energy that accelerates the gas and drives outflows. In addition to these mechanisms, cosmic rays (CRs) provide a form of non-thermal feedback that can further influence gas dynamics and contribute to driving galactic winds \citep{Zweibel2013,Zweibel2017, Ruszkowski_Pfrommer2023}. These relativistic particles predominantly interact with the thermal gas through particle-wave interactions, rather than direct collisions, transferring energy and momentum. This process results in an additional pressure that influences gas dynamics and contributes to outflows. CRs can drive galactic winds, contributing to both \textit{ejective} and \textit{ejective} feedback, and they also regulate star formation rates (SFRs) through \textit{preventive} feedback. By adding pressure and energy to the circumgalactic medium (CGM), they reduce gas accretion onto the interstellar medium (ISM) and limit the fuel available for star formation.

CRs are primarily accelerated via diffusive shock acceleration in supernova remnants and likely also in AGN-powered jets \citep{Krymskii1977, AxfordEtAl1978, Bell1978a, Bell1978b, BlandfordOstriker1978}. Unlike thermal energy, CRs -- especially GeV protons -- are long-lived and can persist for Gyr timescales, allowing them to propagate far from their injection sites. This enables CRs not only to impart momentum to the ISM but also to continually rejuvenate the launched winds through thermal and dynamical coupling of plasma and CRs. 

Because CRs propagate primarily along magnetic field lines, their transport is fundamentally anisotropic, making magnetohydrodynamics (MHD) essential for their modeling. Because of their complex transport mechanisms, CRs are often neglected in cosmological simulations of galaxy formation. While many idealised and cosmological simulations have demonstrated that CRs can significantly affect galaxy evolution -- by driving outflows, regulating star formation, and altering gas properties -- these studies do not yet provide a consistent picture. The predicted impact of CRs varies strongly depending on the assumed implementations of stellar and AGN feedback, the modeling of CR injection and transport (e.g. diffusion coefficients, cooling processes), and the numerical schemes used to evolve CRs. As a result, the role of CRs in galaxy formation remains uncertain.

To better understand these effects, previous works have explored CR feedback in a range of setups. Idealised simulations of the star-forming ISM have shown that CRs can help launch winds and suppress star formation \citep{Simpson2016, Girichidis+2016, Farber+2018, Simpson+2023, Rathjen+2021, Rathjen+2023, Sike+2025}. Similar effects have been reported in simulations of isolated galaxies \citep{Uhlig+2012, Booth+2013, Salem+Bryan2014, Pakmor2016b, Pfrommer+2017, Jacob+2018, Butsky+Quinn2018, Dashyan+Dubois2020, Quataert+2022, Girichidis+2022, Farcy+2022, Thomas+2023, Thomas+2024} and in cosmological contexts \citep{Jubelgas+2008, Salem+2014b, Buck2020, Chan+2019, Hopkins+2021a, Hopkins+2021b, Rodriquez+2023}. However, the predicted impact of CRs on star formation, gas morphology, and outflows varies across studies due to uncertainties in stellar and AGN feedback models, CR injection and transport physics, and their numerical implementation.

Some simulation suites focus on how gas surface density and halo mass influence the role of CRs and their impact on the ISM and CR-enhanced winds in isolated galaxies \citep{Uhlig+2012,Jacob+2018, Rathjen+2023}. These studies find that the effect of CRs on mass loading -- defined as the ratio of the mass outflow rate to the SFR -- decreases with higher surface density and halo mass. Cosmological studies have primarily focused on CR-driven winds in halos of limited mass range, which restricts our understanding of how CRs influence galactic winds across a broader spectrum of halo masses. \citet{Hopkins2019}, however, have examined CR-driven winds across a range of halo masses within the FIRE framework. They find that CRs have a relatively weak effect in low-mass halos ($\Mtwohundred \leq 10^{11}~\Msolar$) and at high redshifts. However, in higher-mass halos ($\Mtwohundred \geq 10^{11}~\Msolar$), CRs can suppress star formation and reduce stellar masses by factors of 2–4 at various redshifts. This occurs in their simulations when the diffusion coefficients are sufficiently high for CRs to escape dense, star-forming regions and build up a CR-pressure-dominated CGM that sustains cool gas and regulates inflows.

Several MHD simulation codes have been developed to study the acceleration and transport of CRs in galaxies, enabling more comprehensive investigations into their effects on galaxy evolution \citep{Miniati+2001, Hanasz+Lesch2003, Pfrommer+2006, Ensslin+2007, Jubelgas+2008, Booth+2013, Salem+Bryan2014, Dubois2016, Pakmor2016Diffusion, Pfrommer_ArepoCR2017, Mignone+2018, Chan+2019, Jiang+Oh2018, Thomas+Pfrommer2019, Thomas2021}. In this paper, we utilise the \textsc{Arepo} code \citep{Arepo, Pakmor2016}, combined with the CR physics implementation described in \citet{Pakmor2016Diffusion} and \citet{Pfrommer_ArepoCR2017}, to investigate how CR-driven winds behave in a cosmological context. We present a series of cosmological zoom simulations of halos ranging from the dwarf scale ($\Mtwohundred = 10^{10}~\Msolar$) to small galaxy groups ($\Mtwohundred = 10^{13}~\Msolar$), extending the work of \citet{Buck2020}, who used the same Auriga galaxy formation model with the same implementation of CR physics to study Milky Way (MW)-like galaxies. The primary aim of this study is to explore how the impact of CRs on galactic winds changes with halo mass in a cosmological context, and how this depends on the chosen CR transport parameters and acceleration efficiencies. In addition, we examine how CRs influence galaxy properties, such as SFRs and gas morphology, as well as the structure and composition of the CGM.

This paper is structured as follows. In Section~\ref{sec:surge}, we introduce the galaxy sample used in this study, which is part of the \textsc{SURGE} (\textbf{S}imulating the \textbf{U}niverse with \textbf{R}efined \textbf{G}alaxy \textbf{E}nvironments) suite. Section~\ref{sec:globalprops} presents an analysis of the global galaxy properties at $z=0$. In Section~\ref{sec:gasprops}, we explore the influence of CRs on gas properties. Section~\ref{sec:sfh} examines how CR feedback modulates SFRs. Section~\ref{sec:outflow_massloading}
analyses the dynamical effects of CRs on outflow properties, focusing on wind acceleration and mass loading. In Section~\ref{sec:CRsVar} we then examine variations in CR physics and explore how different CR transport parameters influence the results. Section~\ref{sec:discussion} provides a discussion of the findings. Finally, in Section~\ref{sec:summary} we summarise our results and provide a brief outlook. 
\begin{table}
\begin{center}
\begin{tabular}{ l c c l l }
\hline
Name & CRs & Alfvén & $\kappa$         & $\zeta_{\text{SN}}$  \\
     &     &  cooling      & [cm$^2$~s$^{-1}$]  &   \\
\hline 
\hline
standard & \xmark\ & \xmark\ &  \xmark & \xmark\  \\
+CR &  \cmark\ & \cmark\ &  $10^{28}$ & 0.1  \\
+CR $-$Alfvén & \cmark\ & \xmark\ &  $10^{28}$ & 0.1  \\
+CR, $\kappa$3e28 & \cmark\ & \cmark\ &  $3\times10^{28}$ & 0.1  \\
+CR, $\kappa$1e29 & \cmark\ & \cmark\ &  $10^{29}$ & 0.1  \\
+CR, $\zeta$05 & \cmark\ & \cmark\ &  $10^{28}$ & 0.05  \\
\hline
\end{tabular}
\end{center}
\caption{Overview of the physical models included in the different simulation runs. Each row lists one of the simulation variants, indicating whether CRs are included, whether Alfvén cooling is active, the value of the anisotropic CR diffusion coefficient $\kappa$, and the CR acceleration efficiency $\zeta_{\mathrm{SN}}$. A check mark indicates that the corresponding physical process is included, and a cross denotes that it is not.}
\label{tab:simNames}
\end{table}
\begin{table*}
\begin{center}
\begin{tabular}{ l l l l l l l l l l l l }
\hline
\hline
Name & $\Mtwohundred$ & $\Rtwohundred$ & $M_*$ & SFR & $r_\mathrm{gas}$ & $r_\mathrm{hmr}$ & 
$r_\mathrm{hlr}$ & $\dot{M}_{\mathrm{out};\, 0.25 \times \Rtwohundred}$ & $\dot{M}_{\mathrm{out};\,\Rtwohundred}$ & $\varv_{\mathrm{out};\, 0.25 \times \Rtwohundred}$ & $\varv_{\mathrm{out};\, \Rtwohundred}$ \\
& $[\mathrm{M}_\odot]$ & [kpc] & $[\mathrm{M}_\odot]$ & $[\mathrm{M}_\odot\,\mathrm{yr}^{-1}]$ & [kpc] & [kpc] & [kpc] & $[\mathrm{M}_\odot\,\mathrm{yr}^{-1}]$ & $[\mathrm{M}_\odot\,\mathrm{yr}^{-1}]$ & $[\mathrm{km}\,\mathrm{s}^{-1}]$ & $[\mathrm{km}\,\mathrm{s}^{-1}]$ \\
\hline 
\hline
1e10\_h12\_s & $10^{9.9}$  & $41$ & $10^{7.7}$   & $2.1 \times 10^{-3}$ & 0.9  & 1.2  & 0.6  & 0.39    & 0.002 & 8.5 & 0.99 \\
1e10\_h8\_s  & $10^{10.1}$ & $49$ & $10^{8.8}$   & $2.7 \times 10^{-2}$  & 3.2  & 1.0  & 1.0  & 0.005 & 0.019  & 5.7 & 3.0  \\
1e10\_h11\_s & $10^{10.4}$ & $60$ & $10^{8.7}$   & $2.1 \times 10^{-3}$  & 5.7  & 2.5  & 2.3  & 1.3     & 0.036 & 12.7 & 3.5  \\
1e10\_h9\_s  & $10^{10.6}$ & $70$ & $10^{8.9}$   & $9.7 \times 10^{-3}$  & 9.1  & 3.9  & 2.8  & 4.2    & 0.064 & 11.9 & 3.6  \\
1e11\_h10\_s & $10^{10.9}$ & $92$ & $10^{9.6}$   & $8.2 \times 10^{-2}$  & 12.6 & 3.7  & 3.0  & 4.0    & 0.37 & 20.9 & 10.3  \\
1e11\_h11\_s & $10^{11.0}$ & $95$ & $10^{9.7}$   & $8.9 \times 10^{-1}$  & 10.1 & 4.2  & 4.1  & 3.2    & 0.54  & 19.5 & 12.8 \\
1e11\_h5\_s  & $10^{11.4}$ & $138$ & $10^{10.1}$ & $6.8 \times 10^{-1}$  & 21.9 & 7.5  & 6.9  & 9.3    & 2.0  & 13.1 & 26.5 \\
1e11\_h4\_s  & $10^{11.5}$ & $138$ & $10^{10.1}$ & $1.4 \times 10^{-1}$  & 31.9 & 7.4  & 4.6  & 6.2    & 3.5  & 17.9 & 25.8  \\
1e12\_h12\_s & $10^{12.0}$ & $217$ & $10^{10.8}$ & $2.2 \times 10^{0}$  & 21.4 & 8.0  & 9.1  & 22.3    & 16.4  & 60.6 & 45.9 \\
1e12\_h5\_s  & $10^{12.1}$ & $223$ & $10^{10.9}$ & $2.1 \times 10^{0}$  & 23.3 & 5.7  & 8.8  & 12.5    & 11.9  & 54.2 & 46.9  \\
1e13\_h3\_s  & $10^{12.5}$ & $311$ & $10^{11.2}$ & $1.6 \times 10^{1}$  & 34.9 & 14.1 & 16.7 & 16.4    & 47.1  & 100.3 & 66.2 \\
1e13\_h8\_s  & $10^{13.0}$ & $459$ & $10^{11.4}$ & $6.0 \times 10^{1}$  & 46.9 & 19.5 & 35.8 & 140.4    & 305.3 & 156.3 & 77.5  \\
1e13\_h7\_s  & $10^{13.1}$ & $478$ & $10^{11.4}$ & $5.7 \times 10^{1}$  & 0.9  & 94.9 & 49.0 & 660.2    & 350.2  & 167.2 & 123.9  \\
\hline
1e10\_h12\_+CR & $10^{9.8}$  & $40$  & $10^{7.5}$  & $1.9 \times 10^{-3}$ & 0.6  & 1.1  & 0.5  & 0.03  & 0.056 & 6.9 & 8.8  \\
1e10\_h8\_+CR  & $10^{10.1}$ & $49$  & $10^{8.5}$  & $2.4 \times 10^{-2}$ & 1.1  & 0.8  & 0.5  & 0.23  & 0.21 & 13.5 & 16.9  \\
1e10\_h11\_+CR & $10^{10.4}$ & $60$  & $10^{8.3}$  & $2.0 \times 10^{-3}$ & 3.7  & 2.2  & 2.0  & 0.2    & 0.21  & 7.3 & 8.1 \\
1e10\_h9\_+CR  & $10^{10.6}$ & $69$  & $10^{8.7}$  & $7.1 \times 10^{-3}$ & 6.9  & 4.3  & 3.6  & 0.57    & 0.24 & 7.8 & 9.5  \\
1e11\_h10\_+CR & $10^{10.9}$ & $92$  & $10^{9.5}$  & $6.7 \times 10^{-2}$ & 7.7  & 3.3  & 3.4  & 1.4    & 0.5 & 12.6 & 16.1  \\
1e11\_h11\_+CR & $10^{11.0}$ & $95$  & $10^{9.6}$  & $1.3 \times 10^{-1}$ & 8.8  & 3.5  & 3.3  & 1.4    & 0.73 & 13.0 & 16.2   \\
1e11\_h5\_+CR  & $10^{11.4}$ & $137$ & $10^{10.1}$ & $5.6 \times 10^{-1}$ & 19.6 & 6.2  & 6.5  & 3.5     & 1.9  & 14.3 & 22.2 \\
1e11\_h4\_+CR  & $10^{11.5}$ & $138$ & $10^{10.0}$ & $1.2 \times 10^{-1}$ & 20.9 & 7.2  & 6.7   & 4.4    & 3.5  & 16.9 & 28.1 \\
1e12\_h12\_+CR & $10^{12.0}$ & $218$ & $10^{10.8}$ & $1.9 \times 10^{0}$ & 15.2 & 5.2  & 6.1   & 23.2    & 11.3  & 63.6 & 31.5 \\
1e12\_h5\_+CR  & $10^{12.1}$ & $224$ & $10^{10.9}$ & $2.0 \times 10^{0}$ & 21.6 & 5.0  & 7.3   & 10.2    & 38.5  & 44.9 & 49.2 \\
1e13\_h3\_+CR  & $10^{12.5}$ & $307$ & $10^{11.2}$ & $5.5 \times 10^{0}$ & 28.8 & 12.4 & 15.0  & 11.4    & 47.1 & 100.3 & 66.2  \\
1e13\_h8\_+CR  & $10^{13.1}$ & $458$ & $10^{11.4}$ & $5.6 \times 10^{1}$ & 19.5 & 16.8& 19.3   & 125.3   & 205.3   & 99.4 & 47.0 \\
1e13\_h7\_+CR  & $10^{13.1}$ & $482$ & $10^{11.4}$ & $4.9 \times 10^{1}$ & 0.5  & 129.4& 16.2  & 381.6    & 442.8  & 174.1 & 166.7 \\
\hline
1e10\_h12\_+CR -Alfvén & $10^{9.8}$  & $40$  & $10^{7.4}$  & $2.1 \times 10^{-3}$ & 3.8  & 1.0  & 0.4  & 0.035 & 0.041 & 7.8 & 9.4  \\
1e10\_h8\_+CR -Alfvén  & $10^{10.1}$ & $48$  & $10^{8.5}$  & $2.5 \times 10^{-2}$ & 0.9  & 0.9  & 0.5  & 0.28 & 0.26 & 15.9 & 26.0  \\
1e10\_h11\_+CR -Alfvén & $10^{10.4}$ & $60$  & $10^{8.1}$  & $1.9 \times 10^{-3}$ & 3.1  & 2.3  & 1.4  & 0.18 & 0.3 & 7.2 & 8.5  \\
1e10\_h9\_+CR -Alfvén  & $10^{10.6}$  & $69$  & $10^{8.5}$ & $9.0 \times 10^{-3}$ & 6.9  & 4.3  & 2.2  & 0.52 & 0.49 & 8.8 & 9.6 \\
1e11\_h10\_+CR -Alfvén & $10^{10.9}$ & $91$  & $10^{9.3}$  & $6.4 \times 10^{-2}$ & 5.3  & 2.4  & 1.6  & 1.4 & 1.4 & 13.9 & 22.3  \\
1e11\_h11\_+CR -Alfvén & $10^{11.0}$ & $94$ & $10^{9.7}$   & $9.8 \times 10^{-2}$ & 3.8  & 1.0  & 0.4  & 1.5 & 1.0 & 12.2 & 16.2  \\
1e11\_h5\_+CR -Alfvén  & $10^{11.4}$ & $137$ & $10^{9.9}$  & $5.6 \times 10^{-1}$ & 11.4 & 5.7  & 7.3  & 3.7 & 2.6 & 10.6 & 23.8 \\
1e11\_h4\_+CR -Alfvén  & $10^{11.5}$ & $138$ & $10^{9.8}$  & $1.2 \times 10^{-1}$ & 13.1 & 6.3  & 6.9  & 4.7 & 3.3 & 13.2 & 16.4   \\
1e12\_h12\_+CR -Alfvén & $10^{12.0}$ & $216$ & $10^{10.7}$ & $2.0 \times 10^{0}$ & 19.0 & 4.5  & 4.6  & 26.9 & 13.3 & 25.6 & 43.2   \\
1e12\_h5\_+CR -Alfvén & $10^{12.1}$ & $224$ & $10^{10.8}$ & $2.0 \times 10^{0}$ & 19.0  & 1.5  & 2.0  & 35.6 & 14.4 & 26.8 & 32.2 \\
1e13\_h8\_+CR -Alfvén & $10^{13.0}$ & $465$ & $10^{11.4}$ & $5.4 \times 10^{1}$ & 23.1  & 13.0 & 54.2 & 85.7 & 359.0 & 182.2 & 102.3 \\
\hline
1e10\_h8\_$\kappa$3e28  & $10^{9.9}$ & $47$ & $10^{8.4}$    & $2.4 \times 10^{-2}$ & 0.8  & 0.9  & 0.5  & 0.087 & 0.1 & 30.2 & 51.7  \\
1e11\_h10\_$\kappa$3e28 & $10^{10.9}$ & $90$ & $10^{9.4}$   & $6.9 \times 10^{-2}$ & 6.7  & 3.5  & 3.2  & 1.4 & 1.3 & 17.8 & 22.3   \\
1e12\_h12\_$\kappa$3e28 & $10^{12.0}$ & $215$ & $10^{10.8}$ & $1.8 \times 10^{0}$ & 18.3 & 5.2  & 5.5  & 17.8 & 11.0 & 67.2 & 53.7 \\
1e12\_h5\_$\kappa$3e28  & $10^{12.1}$ & $224$ & $10^{10.9}$ & $2.0 \times 10^{0}$ & 16.9 & 3.7  & 7.0  & 17.6 & 31.2 & 68.2 & 44.2  \\
\hline
1e12\_h12\_$\kappa$1e29 & $10^{12.0}$ & $214$ & $10^{10.8}$ & $1.8 \times 10^{0}$ & 15.8 & 6.4  & 7.7  & 20.6 & 23.5 & 64.5 & 48.6  \\
\hline
1e10\_h8\_$\zeta$05  & $10^{10.1}$ & $49$ & $10^{8.5}$ & $2.5 \times 10^{-2}$ & 1.2  & 0.8  & 0.5  & 0.22 & 0.18 & 15.7 & 16.3   \\
1e10\_h11\_$\zeta$05 & $10^{10.4}$ & $60$ & $10^{8.3}$ & $2.0 \times 10^{-3}$ & 3.6  & 2.3  & 2.1 & 0.22 & 0.23 & 7.9 & 7.4   \\
1e11\_h10\_$\zeta$05 & $10^{10.9}$ & $92$ & $10^{9.5}$ & $7.2 \times 10^{-1}$ & 8.1  & 3.4  & 3.6 & 0.97 & 0.37 & 12.5 & 13.6    \\
1e12\_h12\_$\zeta$05 & $10^{12.0}$ & $215$ & $10^{10.8}$ & $1.8 \times 10^{0}$ & 21.3  & 7.2  & 7.7 & 15.9 & 11.9 & 54.2 & 46.9    \\
1e12\_h5\_$\zeta$05 & $10^{12.1}$ & $224$ & $10^{10.9}$ & $2.1 \times 10^{0}$ & 15.4  & 4.7  & 7.3 & 13.1 & 30.7 & 75.8 & 48.1    \\
1e13\_h8\_$\zeta$05 & $10^{13}$ & $456$ & $10^{11.4}$ & $5.6 \times 10^{1}$ & 23.0  & 18.6  & 14.9 & 125.3 & 251.6 & 181.3 & 111.0  \\
\hline
\hline
\end{tabular}
\end{center}
\caption{Properties of the simulated galaxies at $z=0$. Columns show, from left to right, the name of the halo (where the s indicates the standard simulation), its total mass $\Mtwohundred$ and radius $\Rtwohundred$, the total stellar mass $M_*$, and SFR averaged over the last $100 \, \mathrm{Myr}$. We also list the gas disc radius $r_\mathrm{gas}$, the stellar half-mass radius $r_\mathrm{hmr}$, and the stellar half-light radius $r_\mathrm{hlr}$. The final columns give the mass outflow rates $\dot{M}_\mathrm{out}$ and outflow velocities $\varv_\mathrm{out}$ measured at $0.25 \times \Rtwohundred$ and $\Rtwohundred$, each averaged over the past $100 \, \mathrm{Myr}$.}
\label{tab:galprops}
\end{table*}

\section{Methods and the SURGE galaxy sample}
\label{sec:surge}

We selected 13 relatively isolated haloes with halo masses\footnote{The virial mass is defined as the mass enclosed within a sphere where the mean matter density is 200 times the critical density, $\rho_\text{crit} = 3H^2(z) / 8\pi G$. Virial quantities are measured at this radius and denoted with a `200c' subscript.} at redshift $z=0$ ranging from $\Mtwohundred = 10^{10}\Msolar$ to $10^{13}\Msolar$. These haloes were drawn from the dark matter-only EAGLE simulation with a comoving side length of 100 Mpc \citep{Schaye2005Eagle} and re-simulated at significantly higher resolution using cosmological MHD zoom-in simulations. The high-resolution region encompasses the Lagrangian volume of all dark matter particles within the halo at $z=0$, extended to $\approx 5 \times R_{200c}$. The initial conditions were created using \textsc{Panphasia} \citep{Jenkins:2013}. We selected two haloes for every 0.5 dex increment in halo mass to enable comparisons within the same mass range. In addition, we deliberately chose disc-dominated central galaxies to allow comparison with simulations of isolated galaxies in collapsing halos \citep{Jacob+2018}, which are typically disc dominated by construction.

The simulations used the \citet{PlanckCollaboration:2014} cosmological parameters $\Omega_\text{M} = 0.307$, $\Omega_\Lambda = 0.693$, $h = 0.6777$, $\sigma_8 = 0.8288$, and $n_s = 0.9611$.

We computed the dwarf galaxies with halo masses of $\Mtwohundred = 10^{10}\,\Msolar$ and $ 10^{10.5}\,\Msolar$ with baryonic element masses of $8 \times 10^2\,\Msolar$ (``level 2'')\footnote{The ``level'' terminology follows the convention introduced in the Aquarius project \citep{Springel2008}, where lower level numbers correspond to higher numerical resolution.}. Intermediate-mass halos, with $\Mtwohundred = 10^{11}~\Msolar$ and $10^{11.5}~\Msolar$, were run at a baryonic mass resolution of $6 \times 10^3~\Msolar$ (``level 3''). Finally, the massive halos with $\Mtwohundred \geq 10^{12}~\Msolar$ were simulated with a baryonic resolution of $5 \times 10^4~\Msolar$ (``level 4''). All simulations use an equal number of dark matter and baryonic resolution elements, with dark matter particle masses approximately six times larger than the baryonic mass resolution.

Our simulations were executed using the \textsc{Arepo} code \citep{Arepo, Pakmor2016, Weinberger2020}. \textsc{Arepo} employs a moving Voronoi mesh for solving ideal MHD and utilises a second-order finite volume scheme \citep{Pakmor2011, Pakmor2013}. Magnetic fields are evolved using the Powell scheme for divergence control \citep{Powell1999}, with a relative divergence error of typically a few per cent \citep{Pakmor2020}. Self-gravity is accounted for using a tree-PM scheme and is coupled to MHD through a second-order Leapfrog scheme \citep{Arepo, Gadget4}.

We use the \textsc{Auriga} model \citep[Auriga;][]{Auriga} to incorporate additional physical processes that are relevant for modelling the formation and evolution of galaxies. These processes include radiative cooling of hydrogen and helium, metal line cooling, and heating by a time-evolving UV background that is spatially uniform \citep{Vogelsberger2013}. We utilise an effective model for the multiphase ISM \citep{Springel2003}, which approximates the complex nature of gas in different phases. The model incorporates a stochastic approach to star formation, accounting for metal production and mass loss from star particles due to stellar winds and supernovae \citep{Vogelsberger2013}. 

Additionally, we include an effective model for galactic winds, used in the Auriga simulations, where isotropic winds are launched stochastically from star-forming regions. Wind particles are ejected with velocities tied to the local dark matter velocity dispersion, ensuring they scale with the gravitational potential. These wind particles are initially decoupled from the gas, only interacting gravitationally with the ambient gas. They are launched from high-density, star-forming regions by construction, and re-couple with the surrounding medium once they encounter a gas cell with density below a pre-determined threshold or exceed a maximum travel time. Upon re-coupling, they deposit their mass, momentum, energy (kinetic and thermal), and metals into the gas cell. The threshold density is set to 0.05 times the physical density threshold for star formation (i.e., $n = 0.05 \times 0.11$~cm$^{-3}$ = 0.0055~cm$^{-3}$). 

We also include a framework for the formation and growth of supermassive black holes, incorporating their feedback through AGN processes. Black holes are seeded with an initial mass of $10^5\,h^{-1}\Msolar$ in haloes with masses greater than $5 \times 10^{10}\,h^{-1}\Msolar$. They grow through mergers and through gas accretion, which is modelled as Eddington-limited Bondi-Hoyle-Lyttleton accretion with an additional term accounting for heating losses from the surrounding hot halo. Feedback is implemented in two modes that operate simultaneously: a quasar mode, where thermal energy is injected isotropically into nearby cells, and a radio mode, where energy is deposited into large-scale bubbles distributed around the black hole. For more information on the Auriga model we refer to the original paper \citep{Auriga} and the accompanying release paper \citep{AurigaRelease}.

The \textsc{Auriga} model has been extensively tested and widely used in studying MW-like galaxies, their CGM, and satellite systems \citep[see, e.g.][]{Auriga,AurigaCGM,Simpson2018,Grand2021}. Building on this foundation, the simulations presented here apply this model to a significantly wider range of halo masses \citep[see also][]{Pakmor2024}. This study examines the dynamical impact of CRs on galaxy evolution, focusing on how their influence on morphology, star formation, and outflows varies with halo mass. 

In all simulations, we introduce a spatially uniform magnetic seed field with a comoving strength of $10^{-14}~\mathrm{G}$ at the start of the simulation at $z=127$. As demonstrated by \citet{Pakmor2014} and \citet{Garaldi2021}, the strength of the seed field and the specific seeding mechanisms have a negligible impact on the properties of magnetic fields in MW-like galaxies and their CGM by redshift $z \le 1.5$, provided the simulations achieve sufficient resolution. In \citet{Pakmor2024} we showed that for the numerical resolution we use in the simulations shown in this paper, the magnetic field evolution is converged. 

Building on this work, we ran the same cosmological zoom simulations as in \citet{Pakmor2024} with the same initial conditions but now incorporating CRs \citep{Pfrommer_ArepoCR2017, Pakmor2016Diffusion, Buck2020}, which are accelerated as part of the stellar feedback process. In this work, we assume that CRs are primarily accelerated in supernova remnants and inject them at the time of star particle formation as part of the stellar feedback model, reflecting the short delay times of core-collapse supernovae. Specifically, the CR energy injected is calculated as the product of the number of supernovae per solar mass of stars formed ($10^{-2}$), the energy released per supernova ($10^{51} \, \mathrm{erg}$), and the CR acceleration efficiency ($\zeta_\mathrm{SN}$). The total CR energy is promptly deposited into the neighbouring cells of the newly formed star particle. This approach assumes that the integrated return of CR energy over the lifetime of the stellar population occurs instantaneously. Note that the CR injection process is distinct from the modeling of supernovae feedback. In our setup, the thermal energy from supernovae feeds into the effective equation of state of the multiphase ISM, following \citet{Springel2003}, while CRs are injected as a separate energy component that does not contribute to this equation of state. Furthermore, wind particles in the simulation do not transport CR energy or the magnetic field.

We set the CR acceleration efficiency to $\zeta_\mathrm{SN} = 0.1$ as the fiducial value and vary it in a few simulations to assess its impact. This choice is informed by hybrid particle-in-cell simulations of collisionless non-relativistic shocks, which suggest a maximum energy injection efficiency of approximately 0.15 for quasi-parallel shocks -- those propagating nearly parallel to the upstream magnetic field direction \citep{Caprioli2014}. Integrating the CR energy injection efficiency over random shock obliquities results in an average efficiency of around 0.05 \citep{Pais2018}. Observations of CRs, radio emission, and gamma-ray data in galaxies support an average CR energy injection efficiency of $\zeta_\mathrm{SN} = 0.05 - 0.1$ \citep{Werhahn2021a, Werhahn2021b, Werhahn2021c, Werhahn2023, Pfrommer+2017, Pfrommer2022}.

Assuming that roughly one core-collapse supernova occurs per $100~\Msolar$ of stars formed, and each supernova releases $10^{51}~\mathrm{erg}$, this corresponds to a CR energy injection of $10^{48}~\mathrm{erg} / \Msolar$ per stellar mass formed for our fiducial choice of $\zeta_\mathrm{SN} = 0.1$. For $\zeta_\mathrm{SN} = 0.05$ and 0.15, this energy input becomes $5 \times 10^{47}$ and $1.5 \times 10^{48}~\mathrm{erg} / \Msolar$, respectively. 

The different physical setups used in our simulations are summarised in Table~\ref{tab:simNames}, including whether CRs, Alfvén cooling, and diffusion are included, as well as the adopted values for the diffusion coefficient and CR acceleration efficiency. The resulting galaxy properties at $z=0$ for each simulation are listed in Table~\ref{tab:galprops}.
\subsection{Model for CR transport}
We describe the evolution of CRs using a fluid approximation that captures their collective transport and interactions with the gas. While individual CRs move with velocities close to the speed of light, their interactions with Alfvén waves nearly isotropise their distribution function in the wave frame, resulting in an observed anisotropy in our Galaxy that is remarkably small ($\mathcal{O}(10^{-4})$) at GeV to TeV energies \citep{Kulsrud2005}. In our models, CRs are treated as a relativistic fluid with an adiabatic index of $\gamma_\mathrm{CR} = 4/3$ within a two-fluid approximation \citep[see][]{Pfrommer_ArepoCR2017}. In this framework, the CR fluid predominantly represents GeV protons, which carry most of the CR energy density and dominate the pressure budget relevant for galaxy-scale dynamics \citep[e.g.][]{Strong+2007,Grenier+2015}. We evolve the energy density of CRs and couple the CR fluid to the thermal gas both dynamically, through CR pressure, and thermally, through CR heating and cooling processes. We use the one-moment method to approximate CR transport, which evolves the isotropic part of the CR distribution. This modeling facilitates the representation of CR energy transport as a superposition of CR advection with the gas and anisotropic transport along magnetic field lines, using an effective transport model that includes both diffusion and one of the main effects of streaming through Alfvén wave cooling. Specifically, the CR energy density, $\epsilon_\mathrm{CR}$, evolves in our simulations according to
\begin{align}
\hspace*{35pt}
    \frac{\partial \epsilon_\mathrm{CR}}{\partial t} &+ \bnabla \bcdot \left[ \epsilon_{\mathrm{CR}} \boldsymbol{\varv} - \kappa \mathbfit{b} (\mathbfit{b} \bcdot \bnabla \epsilon_{\mathrm{CR}})  \right] \nonumber \\
    &= -P_{\mathrm{CR}} \bnabla \bcdot \boldsymbol{\varv} - |\boldsymbol{\varv}_\mathrm{A} \bcdot \bnabla P_\mathrm{CR} | + \Lambda_{\text{CR}} + \Gamma_\mathrm{CR} \; .
    \label{eq:EnergyDensityEvol}
\end{align}
\noindent 
Here $\epsilon_{\text{CR}}$ is the CR energy density, $\boldsymbol{\varv}$ is the gas velocity, $\boldsymbol{\varv}_\mathrm{A}$ is the Alfvén velocity, $\mathbfit{b}$ is the unit vector along the magnetic field, $P_{\text{CR}}$ is the CR pressure, $\kappa$ is the kinetic energy-weighted spatial diffusion coefficient, $\Gamma_\mathrm{CR}$ is the source term for the CR energy, and  $\Lambda_\mathrm{CR}$ represents the loss terms of CR energy from hadronic and Coulomb processes. The Alfvén velocity is defined as $\boldsymbol{\varv}_\mathrm{A} = \mathbfit{B} / \sqrt{4 \pi \rho}$, with $\mathbfit{B}$ the magnetic field and $\rho$ the gas density. 

We assume a spatially and temporally constant diffusion coefficient $\kappa$ along the direction of the magnetic field and no diffusion perpendicular to it \citep[see][]{Pakmor2016Diffusion}. Unless noted otherwise, we use a value of $\kappa = 10^{28}$~cm$^2$~s$^{-1}$ and only vary it for a selection of additional simulations to address the uncertainties of CR transport (see Section~\ref{sec:CRsVar}). Note that diffusion (as well as streaming) is an anisotropic transport process along the mean magnetic field and oriented down the pressure gradient. We do not include CR streaming in our simulations, as this can be more accurately solved with the two-moment method of CR transport (see \citealp{Ruszkowski_Pfrommer2023} for a more detailed discussion and \citealp{Jiang+Oh2018} as well as \citealp{Thomas+Pfrommer2019} and \citealp{Thomas2021} for a derivation and implementation). Unless specified otherwise, we incorporate an anisotropic CR diffusion model that includes the Alfv\'en cooling term (the $| \boldsymbol{\varv}_\mathrm{A} \bcdot \bnabla P_\mathrm{CR} |$ term in Equation~\ref{eq:EnergyDensityEvol}). This term accounts for the energy transfer from CRs to Alfvén waves, which are damped via collisionless processes and thereby heat the surrounding gas \citep[see][]{Buck2020}. These waves are self-excited via the resonant CR streaming instability \citep{Kulsrud1969,Shalaby2023,Lemmerz2024}. The process of CR diffusion conserves CR energy, while accounting for the Alfvén cooling emulates one of the main effects  of CR streaming -- namely, the energy transfer from CRs to the gas via wave damping (\citealt{Wiener2017}, see also the discussion in \citealt{Buck2020}). In Section~\ref{sec:CRsVar} we will discuss how our results depend on this assumption. 

Generally, the Coulomb and hadronic processes are dependent on the energy spectrum of the CRs. We assume an equilibrium momentum distribution for the CRs to model their cooling via Coulomb and hadronic interactions with the ambient gas \citep[as in][]{Ensslin+2007, Jubelgas+2008, Pfrommer_ArepoCR2017}. Note, this is unlike schemes that, in addition to what is described here, also dynamically follow the CR energy spectrum \citep[see][]{Girichidis+2020, Girichidis+2022, Girichidis+2024}. We assume that all Coulomb losses and one-sixth of hadronic losses are thermalised, transferring CR energy into thermal energy \citep{Pfrommer_ArepoCR2017}. The remaining five-sixths of the hadronic losses are assumed to convert directly into $\gamma$-rays and neutrinos, which escape in optically thin conditions such as the ISM.

For a more detailed description and discussion of the approximations used in this paper, we recommend the review by \citet{Ruszkowski_Pfrommer2023}.

Throughout this paper, we refer to simulations without CRs as the \standard{} runs, and to those that include CR transport with anisotropic diffusion and Alfvén cooling as the \CRs{} runs.
\begin{figure*}
    \centering
	\includegraphics[width=\linewidth]{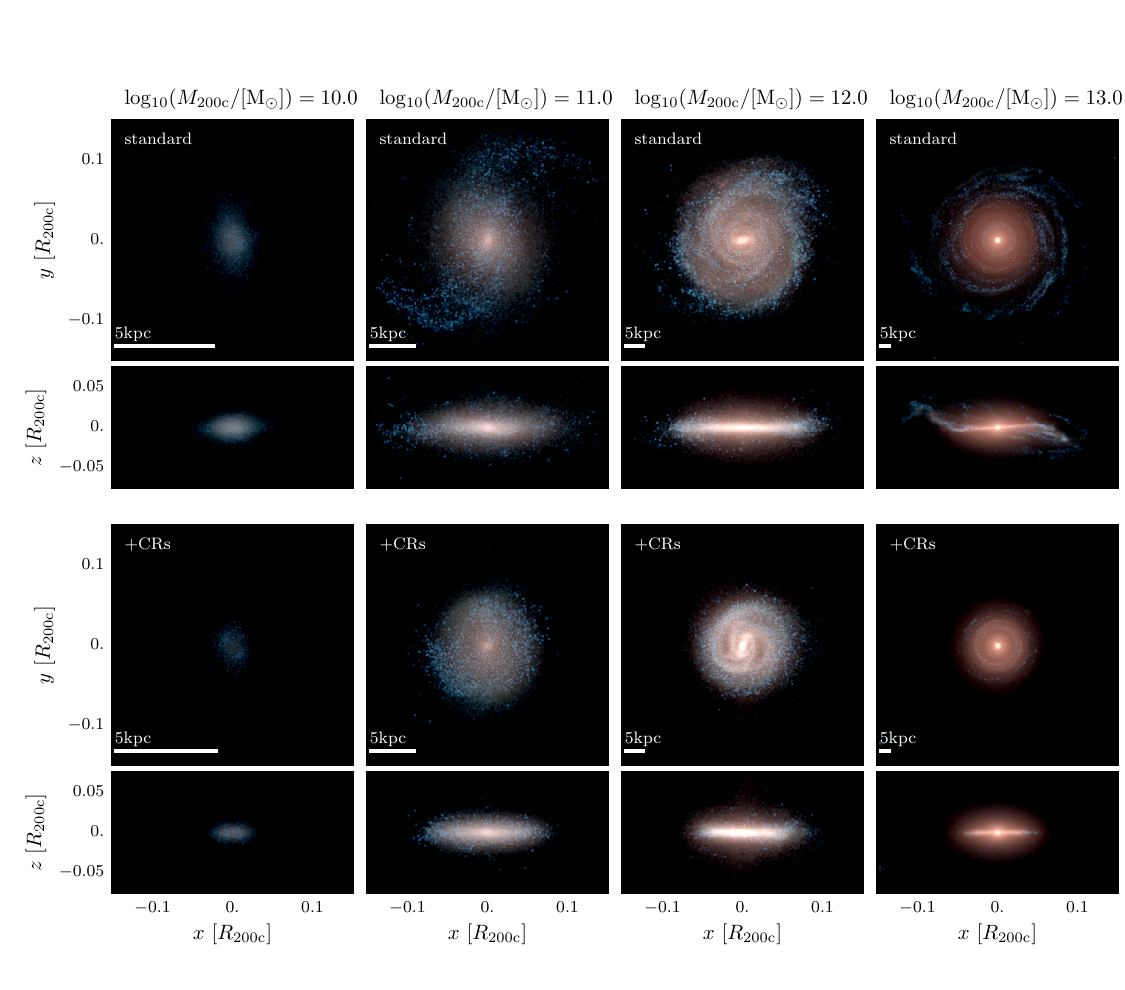}
    \caption{Face-on and edge-on projections of stellar light at $z=0$, with each \standard{} simulation shown on top and the corresponding \CRs{} simulation below. The projection box size is $0.3 \times \Rtwohundred$ in the $x$–$y$ plane and $0.15 \times \Rtwohundred$ along the $z$–axis. The $z$–axis is aligned with the eigenvector of the stellar moment of inertia tensor within $0.1~\Rtwohundred$ that is closest to the total stellar angular momentum vector. We show a representative sample of four galaxies, one for every 1~dex increment in halo mass, to illustrate halo mass dependence. Including CRs makes the stellar disc visually more compact at all halo masses (see also Figure~\ref{fig:overview} for a quantitative assessment).}
    \label{fig:stellarprojection}
\end{figure*}
\begin{figure*}
    \centering
	\includegraphics[width=\linewidth]{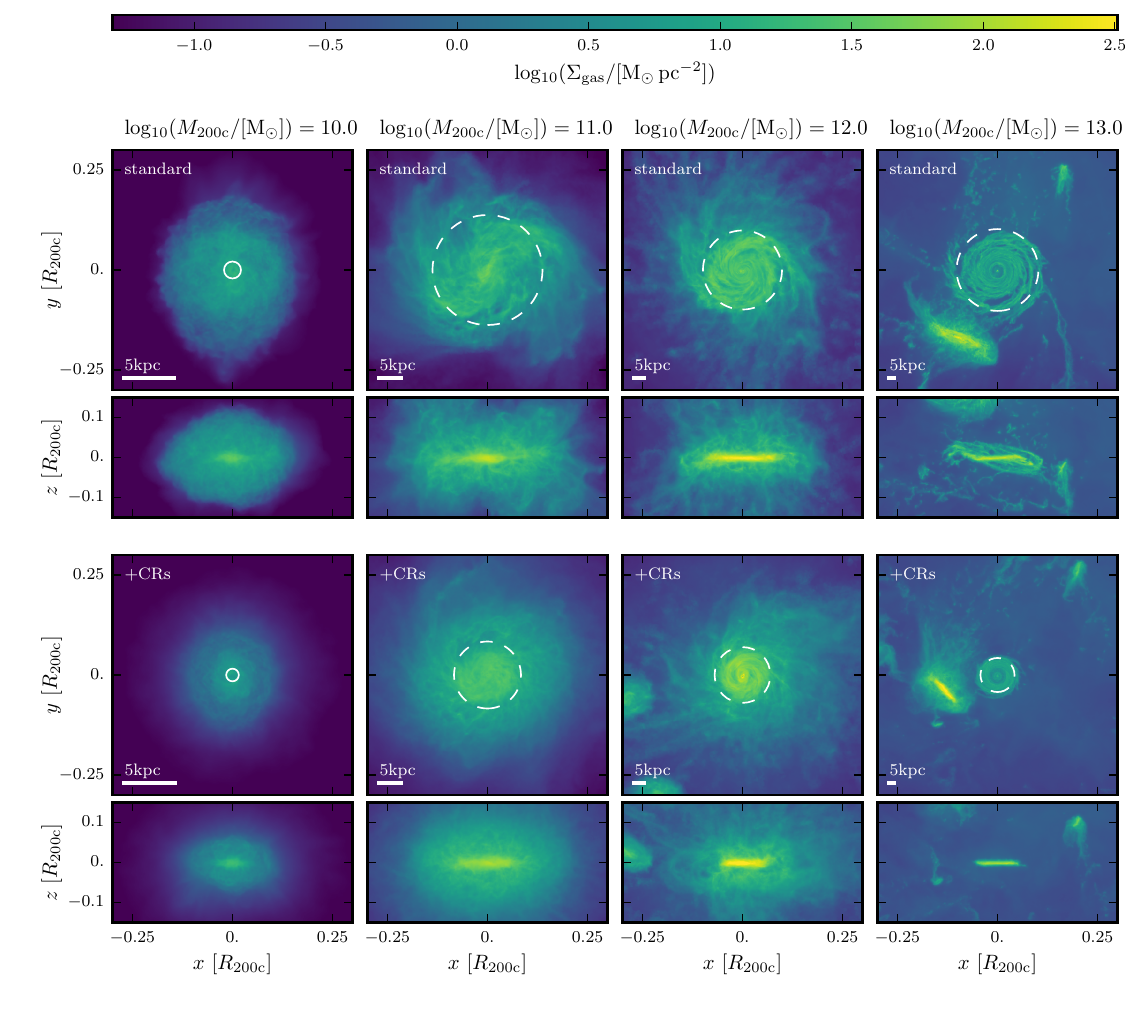}
    \caption{Face-on and edge-on gas surface density maps at $z = 0$, for a representative set of four simulations, shown for both the \standard{} and \CRs{} simulation sets. The size of the projection box is $0.6 \times \Rtwohundred$ in the $x$ and $y$ directions and $0.3 \times \Rtwohundred$ along the line of sight ($z$-axis). The size of the projection box is $0.6 \times \Rtwohundred$. Each \standard{} simulation (top) is paired with its corresponding \CRs{} (bottom) simulation. Dashed circles in the face-on projections indicate the size of the gas disc, defined by an average surface density of $10~\Msolar~\mathrm{pc}^{-2}$. For the lowest mass halo (top/bottom left most column), we apply a cut at $1~\Msolar~\mathrm{pc}^{-2}$ instead. The sizes of the gas disc are visually smaller and more compact across the entire mass range with the inclusion of CRs. Figure~\ref{fig:overview} shows a more quantitative analysis of the size of the gas disc.}
    \label{fig:gasprojection}
\end{figure*}
\begin{figure*}
    \centering
	\includegraphics[width=0.7\linewidth]{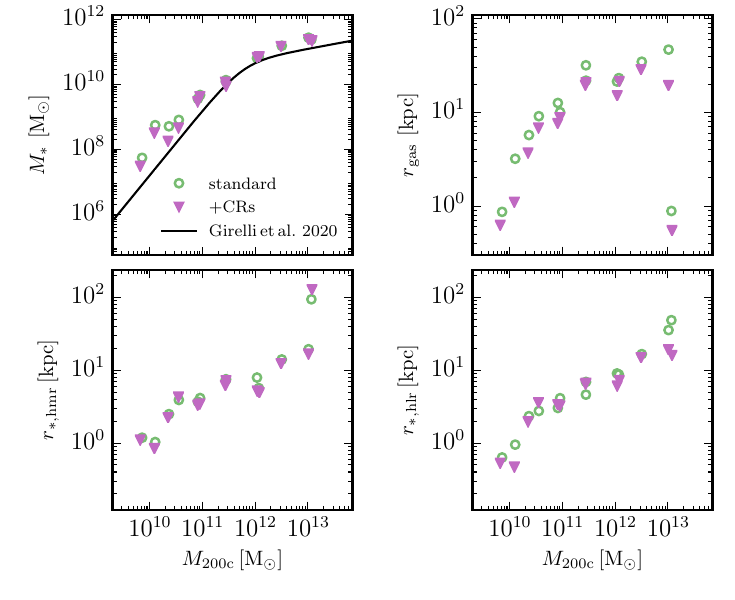}
    \caption{From top left to lower right, the panels show the stellar mass, radius of the gas disc, stellar half-mass radius (hmr), and stellar half-light radius in the $r$ band (hlr) as a function of halo mass for all our cosmological simulations at $z = 0$. The green circles represent simulations conducted with the standard Auriga model, and the pink triangles denote simulations run with CRs. In our simulations, CRs reduce the total stellar mass within the galaxy for halo masses below $\Mtwohundred < 10^{12}~\Msolar$. Additionally, CRs consistently decrease the size of the gas disc across the entire mass range but have no effect on the stellar half-mass radius. CRs generally reduce the half-light radius across the mass range, though the effect is small in several cases and absent in two. The measurements show that CRs influence the gas distribution, leading to smaller gas discs. This, in turn, affects star formation activity, resulting in a smaller half-light radius when CRs are included. However, the half-mass radius remains unchanged because it reflects the overall distribution of stellar mass, which is dominated by older stars formed at earlier times and is less influenced by recent changes in gas distribution or star formation.}
    \label{fig:overview}
\end{figure*}
\section{Global galaxy properties at the present time}
\label{sec:globalprops}
In this Section, we explore how the global properties of galaxies at $z=0$ vary with halo mass and how CRs affect them. 

Figure~\ref{fig:stellarprojection} shows the face-on and edge-on stellar light projections at $z=0$ for a representative sample of galaxies in both the \standard{} and \CRs{} simulations\footnote{The representative galaxies shown here are 1e10\_h12, 1e11\_h10, 1e12\_h12, and 1e13\_h8. These four galaxies (haloes) are used consistently throughout the paper whenever we show only a subset of four galaxies.}. All galaxies, except for the lowest-mass ones, form a stellar disc by $z=0$ in both the \standard{} and \CRs{} runs. Even the smallest galaxy is flattened, as expected, because the sample was selected by eye to include disc-like galaxies for comparison with isolated galaxy simulations. 

Including CRs leads to noticeable changes in the stellar discs. They appear more compact, with fewer blue stars in the outskirts due to suppressed star formation at larger radii. This highlights the influence of CRs on the spatial distribution of young stars and star formation activity (see also Section~\ref{sec:sfh}). 

Figure~\ref{fig:gasprojection} provides a corresponding view of the gas, showing face-on and edge-on projections of gas surface density at $z=0$. The \standard{} and \CRs{} simulations exhibit clear differences in the gas morphology. The inclusion of CRs makes the distribution of gas smoother, a result of the additional CR pressure. CRs also create more centrally concentrated gas discs across all halo masses, as shown by the white circles in Figure~\ref{fig:gasprojection}, which indicate the disc radius defined at an average surface density of $10~\Msolar~\mathrm{pc}^{-2}$ (or $1~\Msolar~\mathrm{pc}^{-2}$ for the lowest-mass halo), consistent with \citet{Pakmor2024}. This trend is also evident in the gas density profiles, which show higher central gas densities in the \CRs{} simulations, as discussed below.

In halos with $\Mtwohundred < 10^{12}~\Msolar$, CRs lead to a redistribution of gas around the disk, resulting in more diffuse gas with a higher proportion at lower densities. This is supported by the mass-weighted probability density function (not shown), which shows an increased amount of gas at lower densities when CRs are included.

It is important to note that the observed difference in gas morphology and ISM properties are influenced by the use of an effective, pressurised equation of state to model the ISM. If we instead modeled a more realistic multiphase ISM and included various wave damping processes, such as ion-neutral damping and non-linear Landau damping, the impact of CRs on the ISM might be reduced, as suggested by recent work \citep{Thomas+2024,Sike+2025}. However, it remains unclear how these results translate to a full cosmological context. Similar findings as ours on the global properties of gas and stellar discs were found by \citet{Buck2020} for MW-like galaxies, which is expected given that they used the same galaxy formation model and that one of our simulated systems is identical to one in their sample.

Having examined the morphological differences, we now focus on a quantitative analysis of the global properties of the simulated galaxies at $z=0$. Figure~\ref{fig:overview} presents the stellar-mass halo-mass relation (first panel), gas disc radius (second panel), stellar half-mass radius (third panel), and stellar half-light radius computed from the projected $r$-band luminosity (fourth panel).

The stellar-mass halo-mass relation is consistent with abundance matching constraints from \citet{Girelli2020} for intermediate-mass halos (i.e., $\Mtwohundred = 10^{11} - 10^{12}~\Msolar$) in both the \CRs{} and \standard{} simulations, as shown in Figure~\ref{fig:overview}. However, as already noted by \citet{Pakmor2024}, the most massive haloes in this set have stellar masses that lie above observed values. This likely reflects, at least in part, limitations in the AGN feedback model, though intrinsic scatter in the stellar–halo mass relation and our by-eye sample selection may also contribute. At the low-mass end, the stellar masses of the galaxies align well with those from the high-resolution LYRA simulations \citep[see][]{Gutcke2022, Pakmor2024}, though they lie somewhat above the extrapolated abundance matching relation from \citet{Girelli2020}. This deviation remains difficult to interpret, given the large scatter in stellar masses found in cosmological simulations at this mass scale \citep{Onorbe2015, Maccio2017, Agertz2020} and the significant uncertainties in the abundance matching method at low halo masses.

Although CRs introduce morphological differences in both stellar and gas properties across all halo masses, they only reduce the total stellar mass in galaxies with halo masses below $\Mtwohundred = 10^{12}~\Msolar$. For the $\Mtwohundred \sim 10^{11}$–$3\times10^{11}\Msolar$ haloes, the difference is small ($\lesssim 0.1$ dex; see Table~\ref{tab:galprops}). This difference arises because CRs significantly reduce the SFR in these lower-mass galaxies at all times, as detailed in Section~\ref{sec:sfh}. 

The second panel of Figure~\ref{fig:overview} shows the sizes of the gas discs at $z=0$, where we define the disc radius as the radius with an average surface density of $10\,\Msolar\,\mathrm{pc}^{-2}$ (or $1\,\Msolar\,\mathrm{pc}^{-2}$ for the lowest-mass halo), consistent with the definition used in \citet{Pakmor2024}. CRs consistently reduce the gas disc radius across all halo masses. This behaviour agrees with the findings of \citet{Buck2020}, who argue that the change in gas disc size is caused by CRs altering the gas flow in the CGM, which subsequently affects the angular momentum distribution of the accreted gas and results in smaller gaseous discs. A detailed investigation of whether this process persists across various mass scales is left for future studies. 

\begin{figure*}
    \centering
	\includegraphics[width=\linewidth]{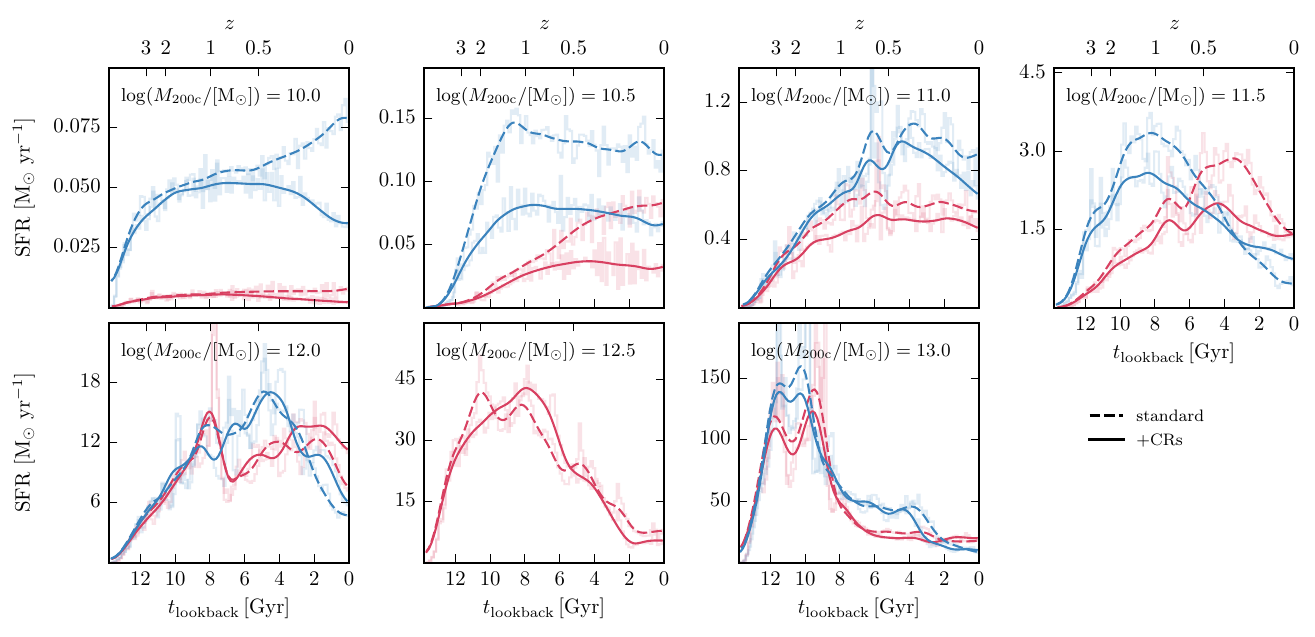}
    \caption{Comparison of star formation histories (SFHs) for different simulations: the \standard{} simulations are represented by a dashed line, while the \CRs{} simulations are shown by a solid line. The faint histograms represent the instantaneous star formation rate, while the overplotted lines, shown more prominently, are smoothed using a Hanning window for clarity. CRs reduce the SFR within haloes with masses less than $10^{12}~\Msolar$. The maximum ratio of SFR between the \standard{} and \CRs{} simulations decreases from a factor of $\sim$~3 at $\Mtwohundred = 10^{10}~\Msolar$  to a factor of $\sim$~1.5 at $10^{11.5}~\Msolar$. For galaxies with halo masses above $\Mtwohundred = 10^{11.5}~\Msolar$, the effect of CRs on SFR diminishes and becomes negligible for the highest-mass galaxies.}
    \label{fig:SFH}
\end{figure*}

The third and fourth panels show the stellar half-mass and stellar half-light radius, respectively. While the stellar half-mass radius shows no clear trend with the inclusion of CRs, the stellar half-light radius tends to be smaller in the presence of CRs (although the effect is small): in 7 out of 13 simulations, it is reduced, in 4 cases it remains largely unchanged, and in 2 cases it is slightly larger. This trend reflects the sensitivity of the half-light radius to the spatial distribution of luminous young and intermediate-age stellar populations, which contribute significantly to the $r$-band emission and are more strongly affected by recent CR feedback. In contrast, the stellar half-mass radius remains largely unchanged, as it traces the full stellar population and is dominated by older stars. The comparison suggests that CRs have a stronger impact at later times, consistent with the star formation histories (SFHs; see Figure~\ref{fig:SFH}), which show a larger suppression in total star formation due to CRs at lower redshifts for most galaxies with $\Mtwohundred < 10^{12}~\Msolar$. A more detailed investigation of the effects of CRs across cosmic time is left for future work.

\textit{\textbf{Summary:}} CRs alter the morphological appearance of galaxies across all halo masses probed in this study. Their impact is quantitatively evident in the global properties of galaxies at $z=0$. Specifically, CRs significantly reduce the stellar masses of our galaxies in halos with masses below $10^{12}~\Msolar$. For higher-mass galaxies (halo masses $\geq 10^{12}~\Msolar$), CRs affect morphology but do not significantly change the total stellar mass. Moreover, CRs consistently reduce the size of the gas disc as well as the stellar half-light radius in most galaxies, irrespective of halo mass. In contrast, the stellar half-mass radius is less affected, as it reflects the cumulative stellar distribution dominated by older stars and is therefore less sensitive to recent CR-driven changes. Since CRs primarily affect recent star formation, their impact is more pronounced in the distribution of younger stars, with limited influence on the total stellar mass profile. Note that, in isolated simulations, such an effect would not be fully captured as they do not capture the full cosmological evolution of galaxies. 
\begin{figure*}
    \centering
	\includegraphics[width=\linewidth]{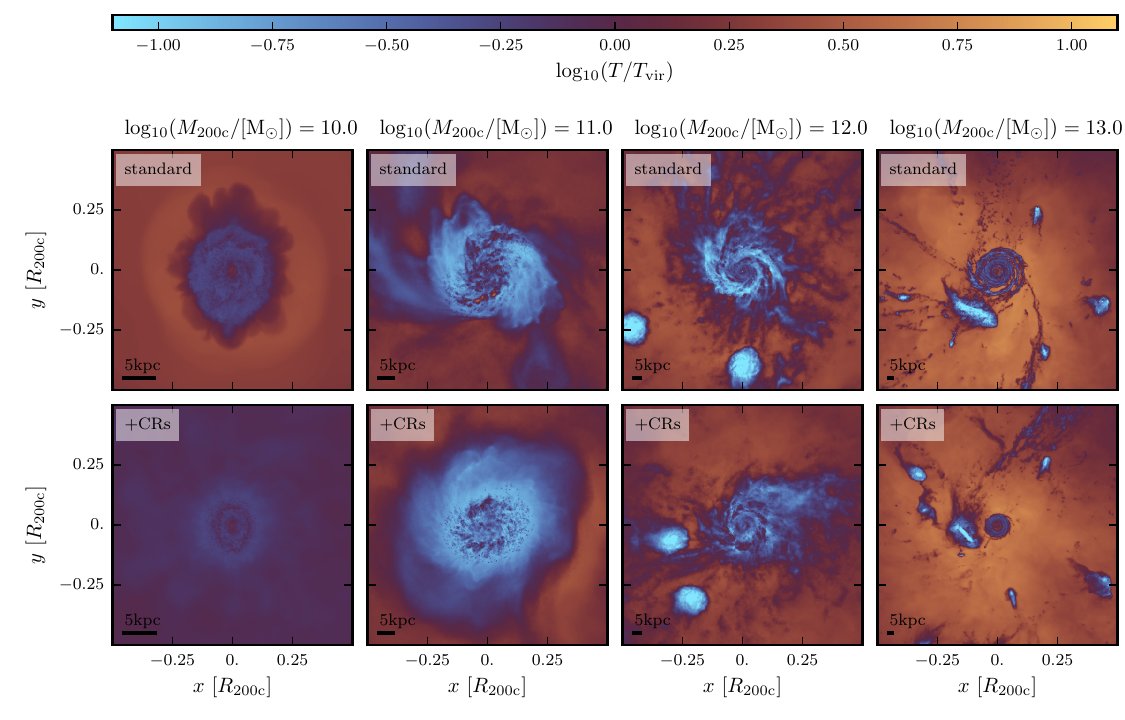}
    \caption{Face-on mass-weighted temperature projections of four representative galaxies, normalised to the virial temperature (i.e, $T_\mathrm{vir} \approx 7.3 \times 10^5\, (M_\mathrm{200c} / 10^{12}~\mathrm{M}_\odot)^{2/3}~\mathrm{K}$) of their halos, shown in 1~dex halo mass increments. The size of the projection box is $1 \times \Rtwohundred$ in both width and depth. The top row shows the \standard{} simulations, while the bottom row displays the \CRs{} simulations. For galaxies with $\Mtwohundred < 10^{12}~\Msolar$, the temperature structure at and around the edge of the galaxy differs noticeably between the \standard{} and \CRs{} simulations. For these halos, the CR simulations reveal cooler surroundings, suggesting that CRs significantly influence the evolution of these lower mass galaxies. In contrast, we find no visible temperature differences in the more massive haloes.}
    \label{fig:tempprojection}
\end{figure*}
\begin{figure*}
    \centering
	\includegraphics[width=\linewidth]{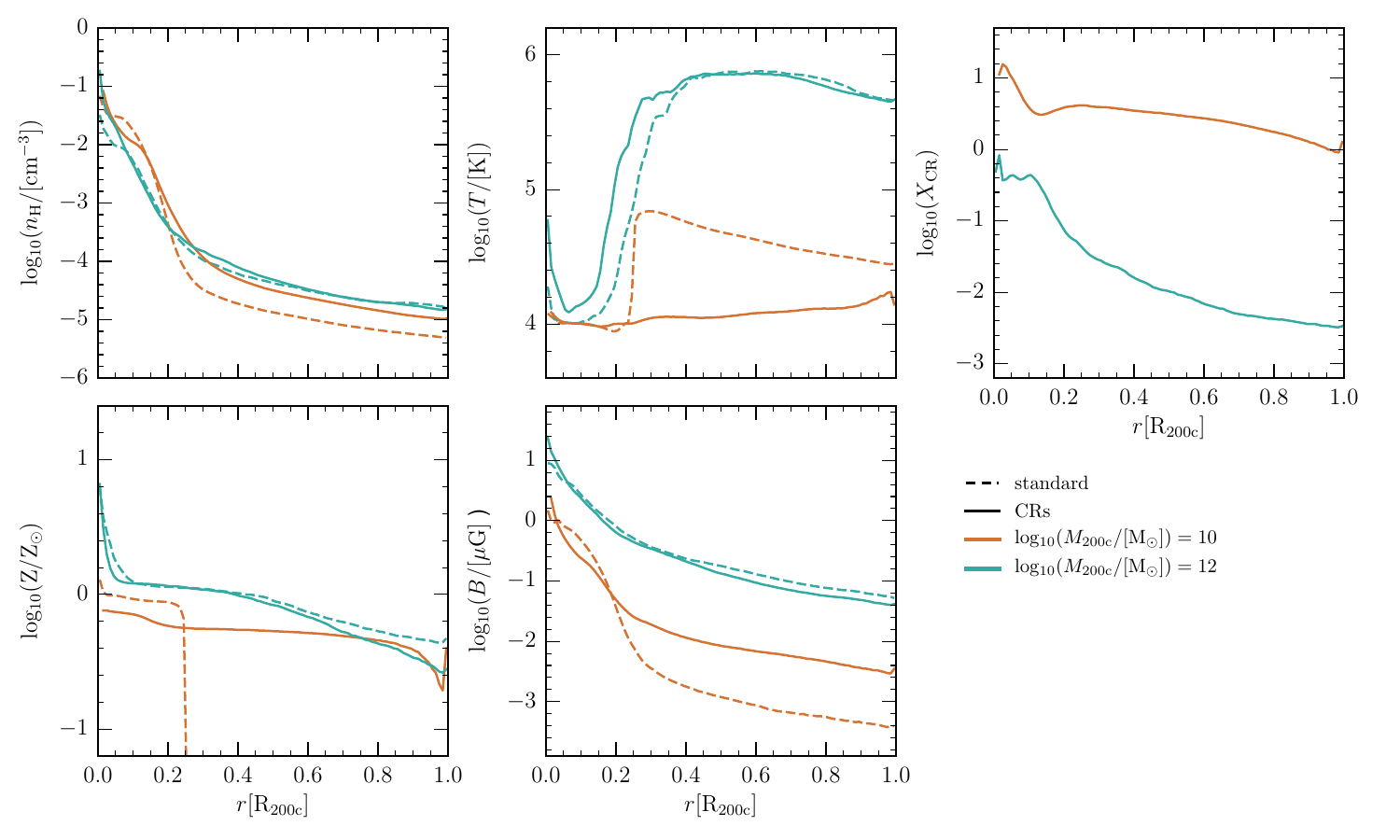}
    \caption{Comparison of radial profiles for two representative simulations with $\Mtwohundred = 10^{10}~\Msolar$ (ochre) and $10^{12}~\Msolar$ (mint), respectively. Each panel shows different profiles for the \standard{} (dashed lines) and \CRs{} (solid lines) simulations. We excluded the satellites and star-forming ISM to calculate all the profiles and averaged over $z=0.1-0$ ($\sim$~1~Gyr). The top row shows, from left to right, the average hydrogen number density profile, the mass-weighted median temperature profile, and the CR-to-thermal pressure ratio profile ($\Xcr =\langle P_\mathrm{CR} \rangle / \langle P_\mathrm{th} \rangle $), calculated as the ratio of the volume-weighted pressure profiles. The bottom row shows the mass-weighted median metallicity profile on the left and the volume-weighted root-mean-square magnetic field strength profile on the right. The inclusion of CRs enhances the density, metallicity and magnetic field strength of the CGM in haloes with masses less than $\Mtwohundred < 10^{12}~\Msolar$. Conversely, the CGM temperature of lower mass galaxies is decreased with the inclusion of CRs. The ratio of CR-to-thermal pressure is above one for the low-mass galaxies within $\Rtwohundred$ whereas for galaxies with masses of $\Mtwohundred \ge 10^{12}~\Msolar$ the ratio is below one. In general, the effect of CRs on the CGM of galaxies with higher masses ($\Mtwohundred \ge 10^{12}~\Msolar$) is small, showing no clear trends.}
    \label{fig:profComp}
\end{figure*}
%
\section{Influence of Cosmic Rays on Gas Properties}
\label{sec:gasprops}

This Section examines the impact of CRs on the gas properties across different halo masses, highlighting how CRs shape the gas distribution and dynamics. 

Figure~\ref{fig:tempprojection} shows face-on temperature projections of four representative galaxies, calculated as the mass-weighted line-of-sight average temperature, normalized to the virial temperature of their halos\footnote{The virial temperature of a halo at redshift zero is defined as $T_{\mathrm{vir}} = \frac{\mu m_\mathrm{p}}{2 k_B} \left( \frac{G M_{200c}}{r_{\text{vir}}} \right)$, where $\mu$ is the mean molecular weight, $m_\mathrm{p}$ the proton mass, $k_B$ the Boltzmann constant, $G$ the gravitational constant. Assuming spherical symmetry, a mean molecular weight $\mu = 0.59$, and the definition of $r_\mathrm{vir}$ as the radius enclosing 200 times the critical density, this yields the approximation: $T_{\mathrm{vir}} \approx 7.3 \times 10^5\, \left( \frac{M_{200c}}{10^{12}~\mathrm{M}_\odot} \right)^{2/3}~\mathrm{K}$.}. For the halo masses shown, ranging from $\Mtwohundred = 10^{10}~\Msolar$ to $10^{13}~\Msolar$, the corresponding virial temperatures are approximately $3.4 \times 10^4$~K, $1.6 \times 10^5$~K, $7.3 \times 10^5$~K, and $3.4 \times 10^6$~K, respectively.

The temperature structure around the edge of the galaxy differs noticeably for galaxies with halo masses of less than $10^{12}~\Msolar$, with significantly lower CGM temperatures when CRs are included. We define the CGM here as all gas beyond the gas disc radius (defined in the previous Section) that is bound to the halo, excluding dense star-forming regions. These findings align with previous studies, which show that including CRs with a substantial transport speed relative to the ambient plasma (e.g.\ an effective CR diffusion coefficient of $\gtrsim10^{28}~\rmn{cm}^2~\rmn{s}^{-1}$) leads to lower plasma temperatures, largely independent of the specific CR transport model, including two-moment implementations \citep{Thomas+2024}. The CR transport speed also determines the exact temperature and which halo masses are most affected. In our simulations, we find that for galaxies with halo masses equal to and above $10^{12}\,\Msolar$, the temperature structure remains largely unchanged between the \standard{} and \CRs{} runs (see Figure~\ref{fig:profComp}). 

To further illustrate the impact of CRs on the gas properties, Figure~\ref{fig:profComp} compares the radially averaged profiles for two representative simulations with a halo mass of $\Mtwohundred = 10^{10}\,\Msolar$ and $10^{12}\,\textbf{}\Msolar$, respectively. These profiles are averaged over 10 snapshots between $z=0.1 - 0$, corresponding to approximately 1~Gyr. We excluded the satellites and cells flagged as star-forming ISM to calculate the profiles. The simulations and halo mass selection shown highlight the varying effects of CRs across different halo masses. 

The density profiles show elevated central values in the CR simulations for both halo masses, with a stronger effect in the higher-mass galaxy. Although we exclude instantaneously star-forming gas from the profiles, some dense, non–star-forming gas remains, and the additional CR pressure supports this gas against gravity, allowing it to reach higher central densities than in the runs without CRs. Beyond approximately 0.2\,$R_{200\mathrm{c}}$, however, the CR runs show enhanced gas densities in the lower-mass galaxies, with little impact observed for the higher-mass haloes. In the lower-mass systems, galaxies in the CR simulations have lower stellar masses than in the standard runs (see Section~\ref{sec:sfh}) but halos with higher baryon fractions in the CGM. This indicates that CRs also act as a form of ``preventive'' feedback, suppressing star formation without fully expelling the gas from the halo and instead allowing more baryons to remain in the CGM where they are supported by CR pressure. In this sense, CRs provide preventive rather than ``ejective'' feedback, as emphasised by \citet{Carr+2023}, although in their case the effect is produced by high energy-loaded, low-density, hot winds. 

The temperature profiles confirm that CRs significantly reduce the temperature of the CGM gas for halos with $\Mtwohundred < 10^{12}~\Msolar$, as already visually apparent in Figure~\ref{fig:tempprojection}. Additionally, the total pressure in the CGM (not shown) is higher in the lower-mass galaxies when CRs are included, driven by an increase in CR pressure, while the thermal pressure remains unchanged. Because part of the pressure support now comes from CRs, the gas can cool to lower temperatures while the system remains still close to hydrostatic equilibrium. The additional CR pressure thus helps to support the denser CGM gas against gravity\footnote{The unchanged thermal pressure despite lower temperatures implies higher densities in the CR runs (see density profiles), but this is less clear in the projection maps since density maps are volume-weighted and temperature maps are mass-weighted.}.

The metallicity profiles show that CRs significantly increase the CGM metallicity in our lower-mass galaxies. This result stems from the stronger outflows driven by CRs (see Section~\ref{sec:outflow_massloading}), which transport metals more effectively, despite the reduction in SFR in these galaxies (see Section~\ref{sec:sfh}). The higher CGM metallicity increases cooling rates, leading to an overall cooler CGM. Furthermore, the additional CR pressure helps to maintain hydrostatic equilibrium in this cooler and denser environment. In contrast, for the higher mass galaxies with halo masses of $\Mtwohundred = 10^{12}~\Msolar$ and above, the metallicity is comparable between the \standard{} and \CRs{} simulations. The smaller changes in metallicity and the lower CR pressure contribution are consistent with the weaker impact of CRs on CGM temperatures in these systems.

This interpretation is supported by the $\Xcr$ profiles, which trace the ratio of CR to thermal pressure ($\Xcr =\langle P_\mathrm{CR} \rangle / \langle P_\mathrm{th} \rangle$). We compute $\Xcr$ as the ratio of the volume-weighted CR pressure profile to the volume-weighted thermal pressure profile. In lower-mass haloes, CR pressure dominates over thermal pressure throughout the CGM, with $\Xcr$ values generally between 1.5 and 4. In contrast, in MW-like haloes, thermal pressure dominates throughout the CGM, with $\Xcr$ values between 0.1 and 0.003\footnote{Note that, unlike our approach, in \citet{Buck2020}, the $\Xcr$ profiles are calculated as the volume-weighted average of $\Xcr$. As a result of this different calculation method, \citet{Buck2020} find $\Xcr$ values above one in the inner CGM even for the MW-like galaxy.}. These values are at least one to two orders of magnitude lower than those found in lower-mass systems. These results demonstrate that CR pressure provides significant support to the CGM in low-mass galaxies, while its contribution remains negligible in more massive haloes in our simulations.

In addition to CR and thermal pressure, magnetic field pressure also contributes to the total pressure budget. In the low-mass galaxy, magnetic pressure is comparable to the thermal pressure in the disc and about 10\% lower in the CGM of the \CRs{} run, while it remains well below the thermal pressure in both regions in the \standard{} run. In the MW-like galaxy, magnetic field pressure is comparable to the thermal pressure in the central regions and stays about 10\% lower at larger radii, in both the \CRs{} and \standard{} runs. These results are consistent with previous findings \citep[e.g.][]{Pakmor2024}, highlighting that magnetic fields can provide a pressure contribution comparable to the thermal component, particularly in more massive systems.

The magnetic field profiles show that CRs increase the magnetic field strength in the CGM of lower-mass galaxies beyond $\sim0.2 \times R_{200\mathrm{c}}$. In contrast, we find no difference in magnetic field strength between the \CRs{} and \standard{} simulations for the more massive galaxies. Since magnetic field amplification in the CGM can result from both turbulence and the transport of magnetised gas by galactic winds, both processes could contribute to the magnetic field strength. The increased magnetic field strength in the lower-mass haloes suggests that CR-enhanced winds manage to push the magnetic field more effectively into the CGM. In these haloes, turbulence is expected to be weak and therefore insufficient to significantly amplify the magnetic field, making outflows the dominant mechanism for CGM magnetisation.

Comparing profiles between all simulations (not all shown) shows that the influence of CRs on gas properties in and around the galaxy systematically decreases towards $\Mtwohundred = 10^{12}\,\Msolar$ and becomes negligible in more massive halos.

\textit{\textbf{Summary:}} Our analysis demonstrates that CRs significantly influence the temperature, density, metallicity, and magnetic field strength of the CGM in our galaxies with halo masses below $10^{12}\,\Msolar$. CRs contribute to denser, more compact galactic gas discs and a CGM characterized by higher metallicity and lower temperatures, primarily through CR-enhanced galactic outflows. These outflows transport metals and push the magnetic field into the CGM. In these low-mass haloes, where turbulence remains weak, outflows act as the primary mechanism for magnetising the CGM. 

It is important to note that these effects are particularly pronounced in the Auriga model, where winds in the standard model are inherently slow in low-mass halos. This arises because the wind properties are tied to the local dark matter velocity dispersion, resulting in slower wind velocities that are less effective at expelling gas from the galaxies. Consequently, the relative impact of CRs may differ in combination with other feedback models. Nonetheless, our findings underscore the critical role of CRs in shaping the physical properties of both the galactic disc and the CGM in lower-mass galaxies.
\begin{figure*}
    \centering
	\includegraphics[width=\linewidth]{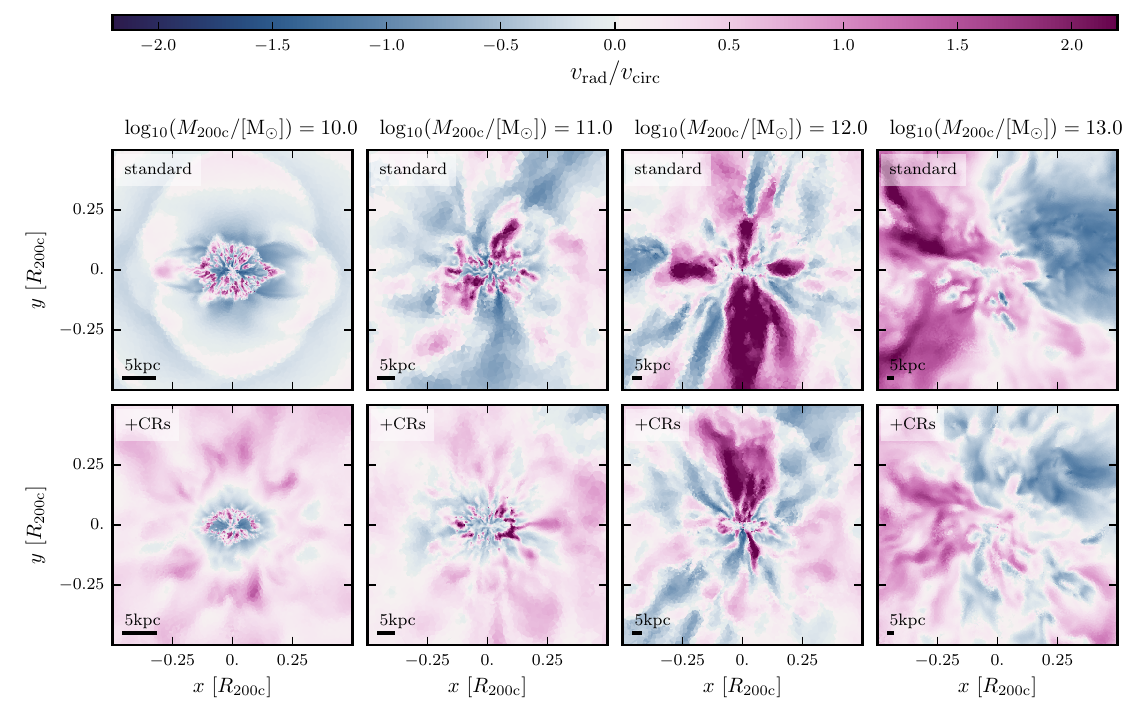}
    \caption{Edge-on radial velocity slices of four representative galaxies, normalised to the circular velocity at half the virial radius of their halos (i.e., $\varv_\mathrm{circ}(0.5\, \Rtwohundred) \approx 164 \, (\Mtwohundred / 10^{12}~\mathrm{M}_\odot)^{1/3}~\mathrm{km\,s^{-1}}$), chosen with 1~dex halo mass increments. The upper row illustrates the \standard{} simulations, while the lower row shows \CRs{} simulations. Differences induced by CRs become apparent in galaxies with $\Mtwohundred < 10^{12}~\Msolar$, with the \CRs{} simulations showing a notable increase in radial velocity in the haloes around these low-mass galaxies.}
    \label{fig:velslice}
\end{figure*}
%
%
\section{Effect of Cosmic Rays on Star Formation}
\label{sec:sfh}

In this Section, we examine the impact of CRs on the SFHs of the cosmological simulations by comparing the \standard{} and \CRs{} simulation sets. Figure~\ref{fig:SFH} shows the SFHs for all simulations, with dashed lines representing the \standard{} simulations and solid lines representing the simulations including CRs. 

CRs reduce the SFR within galaxies with halo masses below $10^{12}\,\Msolar$. The effect of CRs is most pronounced in the lowest-mass galaxies of our sample, where the maximum SFR with CRs is about half of the SFR without CRs. For galaxies with halo masses below $\Mtwohundred = 10^{11.5}\,\Msolar$, the deviation in SFR due to CRs begins early in their evolution, becoming noticeable at high redshifts ($z \sim 3$). For galaxies with halo masses below $10^{11}\,\Msolar$, the suppression of SFR becomes increasingly pronounced over time. In contrast, for galaxies with $\Mtwohundred = 10^{11.5}\,\Msolar$, this suppression weakens or even reverses at low redshift. For the most massive galaxies in our sample ($\Mtwohundred \geq 10^{12}\,\Msolar$), CRs have little to no effect on the SFR both at early and late times.

Even within the same mass range, individual halos exhibit variations in their SFRs at different times due to a combination of halo-to-halo scatter and the stochastic nature of star formation and feedback. To assess the effect of CRs, we compare simulations of galaxies in the same halo mass bin run with and without CRs. In lower-mass halos ($\Mtwohundred \leq 10^{11}\,\Msolar$), the intrinsic variability in total SFR between different halos can exceed the CR-induced suppression seen in matched pairs, highlighting the importance of controlling for halo-to-halo differences in these low-mass galaxies when interpreting the impact of CRs. In addition, stochastic fluctuations in star formation introduce further scatter, even between simulations that start from the same halo initial conditions \citep[see][]{Pakmor+2025}. Quantifying the exact extent of this stochasticity will require future simulations with multiple realisations per halo. 

The reduction in SFRs due to CRs also varies with time, showing different suppression patterns across halo masses. Lower-mass halos ($\Mtwohundred \sim 10^{10}\,\Msolar$) tend to exhibit more consistent reductions in SFR due to CRs, with less temporal variability. In contrast, for more massive halos of $10^{12}~\Msolar$ and above, stochastic behaviour dominates, resulting in relatively small differences between runs of the same halo.

\textit{\textbf{Summary:}} In our simulations, CRs play a significant role in shaping star formation, particularly in galaxies with halo masses below $\Mtwohundred = 10^{12}\,\Msolar$. The impact is most pronounced in the lowest mass haloes (i.e. $10^{10}\,\Msolar$ and $10^{10.5}~\Msolar$), where CRs can suppress the SFR by up to a factor of two compared to simulations without CRs. The influence of CRs on the SFR diminishes towards higher halo masses (i.e., $\Mtwohundred > 10^{12}\,\Msolar$). In intermediate-mass halos (i.e., $10^{11}\,\Msolar$ to $10^{11.5}\,\Msolar$), the reduction in the SFR due to CRs varies significantly in extent and timing, likely reflecting stochastic variations in star formation and feedback.
\begin{figure*}
    \centering
	\includegraphics[width=\linewidth]{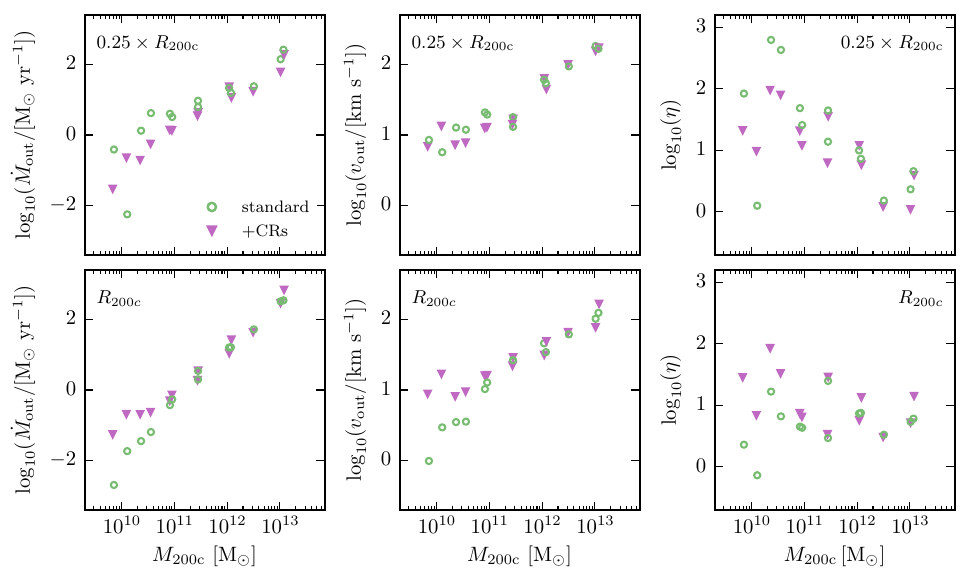}
\caption{Mass outflow rate (left panel; calculated as described in Equation~\ref{eq:masslfux}), radial outflow velocity (middle panel), and mass loading factor (right panel) as a function of halo mass for all cosmological simulations. Green circles and violet triangles show results for the \standard{} and \CRs{} runs, respectively. The top and bottom rows display measurements at $0.25 \times \Rtwohundred$ and $1 \times \Rtwohundred$, averaged over the final $\sim1$~Gyr ($z=0.1$ to $z=0$). 
Near the galaxy, the \standard{} simulations exhibit higher mass outflow rates and mass loading for galaxies with halo masses below $10^{12}\Msolar$, and higher outflow velocities for halos below $\Mtwohundred < 10^{11.5}\Msolar$. In contrast, at the virial radius, the CR runs show -- compared to the \standard{} simulations -- increased outflow rates, velocities, and mass loading for halos with $\Mtwohundred < 10^{11.5}~\Msolar$, indicating that CRs sustain winds over larger distances in lower-mass haloes.}
\label{fig:OutflowMassLoading}
\end{figure*}
%
%
\section{Dynamical Effect of Cosmic Rays on Outflow properties}
\label{sec:outflow_massloading}
In this Section, we examine the influence of CRs on the properties of galactic outflows, which play a pivotal role in regulating star formation by reducing the gas available for star formation. All our simulations include an effective model for galactic winds that drives outflows from galaxies. By comparing outflow measurements from the \standard{} simulations to those that additionally include CRs across all halo masses, we look at the differential impact CRs have on gas dynamics.

Figure~\ref{fig:velslice} shows edge-on radial velocity slices for four representative galaxies, with velocities normalised to the circular velocity at half the virial radius\footnote{The circular velocity at radius $r$ is defined as $\varv_\mathrm{circ}(r) = \sqrt{G M(<\!r)/r}$, where $G$ is the gravitational constant and $M(<\!r)$ is the mass enclosed within $r$. Assuming an NFW halo with concentration $c = 10$, the enclosed mass within $0.5\,R_{200\mathrm{c}}$ is approximately $0.644\,\Mtwohundred$. This gives the approximation: $\varv_\mathrm{circ}(0.5\, \Rtwohundred) \approx 1.135 \, \varv_{200} \approx 164 \left( \Mtwohundred / 10^{12}~\mathrm{M}_\odot \right)^{1/3}~\mathrm{km}~\mathrm{s}^{-1}$.}; the \standard{} simulations are shown in the top row and the \CRs{} simulations in the bottom row. For the halo masses shown, ranging from $\Mtwohundred = 10^{10}$ to $10^{13}~\Msolar$, the corresponding circular velocities at $0.5 \times \Rtwohundred$ are approximately $35$~km~s$^{-1}$, $76$~km~s$^{-1}$, $164$~km~s$^{-1}$, and $353$~km~s$^{-1}$, respectively. For galaxies with halo masses below $\Mtwohundred = 10^{12}~\Msolar$, the radial velocity outside the galaxies shifts from inflow in the \standard{} simulations to outflow in the \CRs{} simulations, highlighting the impact of CRs on the gas dynamics and, subsequently, the evolution of these lower-mass galaxies. In contrast, for higher-mass galaxies ($\Mtwohundred \geq 10^{12}~\Msolar$), we see no visual (nor quantitative) differences in radial velocity in the CGM between the two simulation sets. 

Figure~\ref{fig:OutflowMassLoading} provides a more quantitative comparison of galactic outflow properties across the full range of halo masses ($\Mtwohundred = 10^{10} - 10^{13}~\Msolar$) for the \standard{} and \CRs{} simulations. Outflow measurements are shown at two radial distances: $0.25 \times \Rtwohundred$ and $1 \times \Rtwohundred$, corresponding to the galaxy's immediate vicinity and its virial boundary, respectively. This analysis at different distances helps to better understand how CRs affect outflows both close to and far from the galaxy. All the quantities shown in Figure~\ref{fig:OutflowMassLoading} are averaged between $z=0.1 - 0$ ($\approx 1$~Gyr, which covers 10 simulation outputs). The properties compared include mass outflow rates (first column), radial outflow velocity (second column), and mass loading factors (third column), computed using only gas cells with positive radial velocities.
\begin{figure*}
    \centering
	\includegraphics[width=\linewidth]{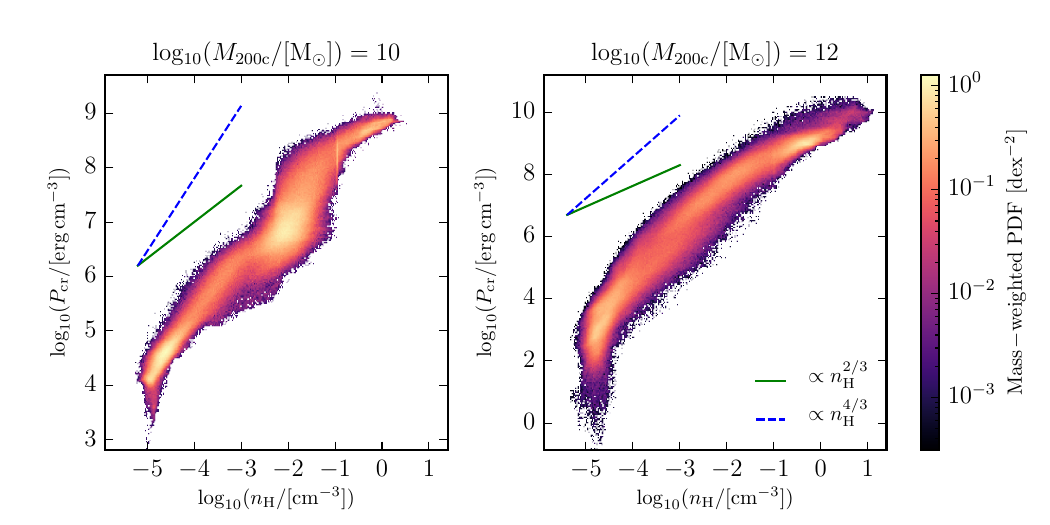}
\caption{Phase diagram of CR pressure versus gas density for the low-mass halo with $\Mtwohundred = 10^{10}~\Msolar$ (left) and the higher-mass halo with $\Mtwohundred = 10^{12}~\Msolar$ (right). The diagrams are colored by the mass-weighted probability distribution function of the gas, excluding gas bound to satellite galaxies. The reference scalings expected for adiabatic evolution ($P_{\rm cr} \propto \rho^{4/3}$) and for pure CR streaming at the Alfvén speed ($P_{\rm cr} \propto \rho^{2/3}$) are shown for comparison. At low densities the distributions in both halos lie between these theoretical limits and tend to follow the adiabatic expectation more closely, while at higher densities the relation becomes more complex due to additional physical processes and the influence of the cosmological environment.}
\label{fig:PcrRho}
\end{figure*}

We compute the mass outflow rate by summing over all outflowing gas cells in a thin spherical shell around the galaxy. The mass flux $\dot{M}_{\text{out}}$ is calculated as:
\begin{equation}
\label{eq:masslfux}
\dot{M}_{\rm out} = \oiint\limits_{\boldsymbol{\varv} \cdot \hat{\mathbfit{r}} > 0} \rho\, (\boldsymbol{\varv} \cdot \hat{\mathbfit{r}})\, \mathrm{d}S \approx \sum_{\substack{i \in \text{shell} \\ \boldsymbol{\varv}_i \cdot \hat{\mathbfit{r}}_i > 0}} m_i \, (\boldsymbol{\varv}_{i} \cdot \hat{\mathbfit{r}}_i) \times \frac{S_{\text{shell}}}{V_{\text{shell}}} \quad .
\end{equation}
Here, $i$ denotes the index of a cell centre within the shell of thickness\footnote{We verified that the results remain quantitatively unchanged when increasing the shell's thickness. Additionally, we checked that the inner radius is outside the recoupling radius of the wind particles.} $\Rtwohundred / 100$. In addition, $m_i$ is the mass, $\boldsymbol{\varv}_{i}$ is the gas velocity, and $\hat{\mathbfit{r}}_i$ is the radial unit vector at the position of the cell centre. $S_{\text{shell}}$ and $V_{\text{shell}}$ are the surface area and volume  of the spherical shell, respectively. To focus on local dynamics relative to the galaxy, we exclude gas within satellites and account for the centre of mass velocity of the halo. Additionally, we calculate the time-averaged mass loading factor as $\langle \eta \rangle = \langle \dot{M}_{\mathrm{out}} \rangle / \langle \mathrm{SFR} \rangle $, where both the outflow rate and the SFR are averaged over the last $\sim$1~Gyr (i.e. ten simulation outputs). 
%
\begin{figure*}
    \centering
	\includegraphics[width=\linewidth]{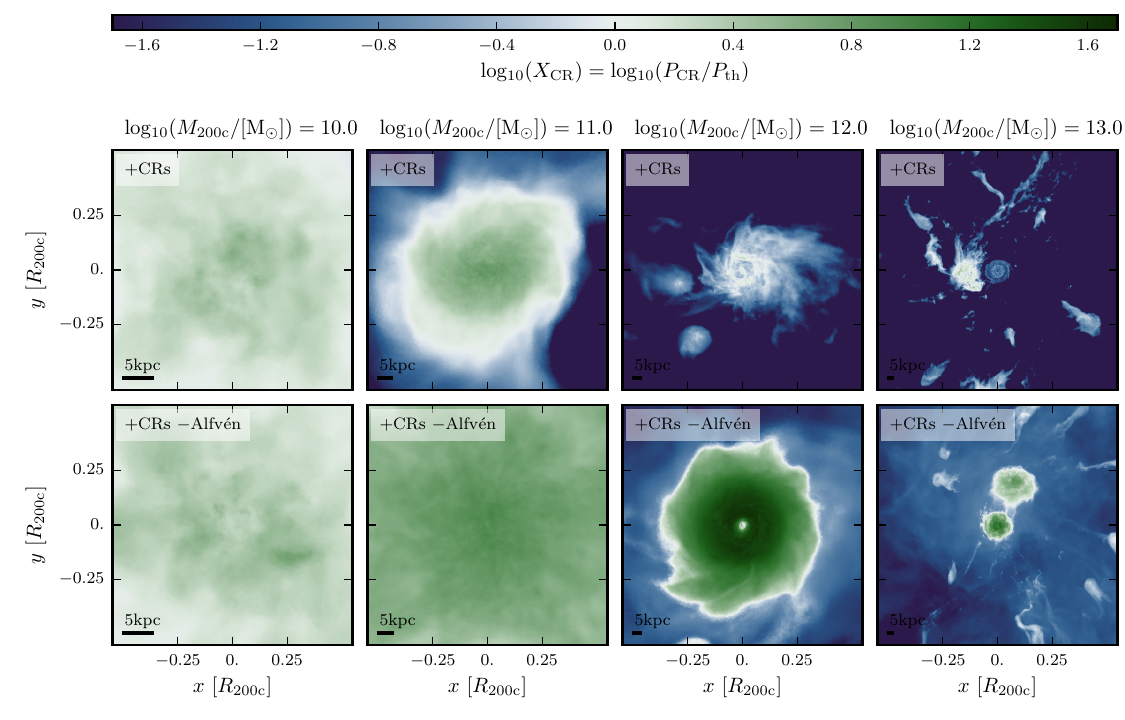}
    \caption{CR-to-thermal pressure ratio ($\Xcr =\langle P_\mathrm{CR} \rangle / \langle P_\mathrm{th} \rangle $) for the fiducial \CRs{} simulation (top) and the simulation excluding the Alfvén cooling term (bottom), shown as a projection through a $\Rtwohundred$-tick slab centered on the galaxy. In the \noalfven{} simulations, CR pressure dominates over thermal pressure within the galaxy and at the disc-halo interface for all galaxies, regardless of halo mass. Notably, for galaxies with halo masses of $10^{12}~\Msolar$ and higher, the CR pressure in the \noalfven{} simulations switches from being comparable to or less than the thermal pressure in the fiducial CR simulations to becoming dominant. This higher-than-unity $\Xcr$ region extends further for MW-like galaxies compared to higher-mass galaxies. In the extended CGM, $\Xcr$ is only slightly increased in the \noalfven{} simulations compared to the fiducial ones for galaxies with halo masses equal to or higher than MW-like galaxies. In lower mass galaxies (i.e., $10^{11}~\Msolar$ and below), CR pressure dominates over thermal pressure within the galaxy and its vicinity (region within $0.25 \times R_{200}$) in both fiducial and \noalfven{} simulations. While minimal differences are observed for the lowest mass galaxy ($10^{10}~\Msolar$), the galaxy with halo mass of  $10^{11}~\Msolar$ shows somewhat higher $\Xcr$ within this region. Outside $0.25 \times \Rtwohundred$, however, $\Xcr$ is strongly increased for $10^{11}~\Msolar$ halos, and marginally for the lowest mass galaxy, when Alfvén cooling is excluded.}
    \label{fig:Xcr}
\end{figure*}
%

In the galaxy's immediate vicinity (i.e., $0.25 \times \Rtwohundred$), the majority of the \standard{} simulations exhibit higher mass outflow rates and mass loading factors than the \CRs{} simulations for galaxies with halo masses below $\Mtwohundred = 10^{12}~\Msolar$. For most galaxies with $\Mtwohundred < 10^{11.5}~\Msolar$, the \standard{} simulations show higher outflow velocities, with the exception of one low-mass `outlier' at $\Mtwohundred = 10^{10}~\Msolar$ discussed more below. Note that the increase in outflow rate is larger than the typical 1$\sigma$ scatter across the 10 snapshots used for averaging. In contrast, the difference in mass loading remains comparable to this scatter due to the large fluctuations in the SFR. For visual clarity, we do not show the 1$\sigma$ scatter in Figure~\ref{fig:OutflowMassLoading}. The higher mass outflow rates and velocities near the galaxy in the \standard{} simulations are likely driven by a combination of factors. The increased SFR in the \standard{} simulations around redshift $z=0$ (see Figure~\ref{fig:SFH}) naturally leads to stronger stellar feedback, ejecting greater amounts of mass, energy, and momentum from the ISM. Note that the combination of variations in the SFH, galaxy morphology, and CGM properties complicate direct comparisons, as these differences would impact outflow properties even in the absence of CRs. 

Interestingly, at the virial radius, the \CRs{} simulations display distinctly higher outflow rates, outflow velocities, and mass loading factors for galaxies with halo masses up to $\Mtwohundred = 10^{11.5}~\Msolar$, while CR effects are negligible for more massive galaxies for the CR transport parameters and galaxy model chosen. As we saw from the top panels, this behaviour contrasts with the trends observed near the galaxy (i.e., $0.25 \times \Rtwohundred$). For these lower-mass galaxies, outflow measurements begin to increase in the \CRs{} simulations compared to the \standard{} simulations at radii around $0.4 \times \Rtwohundred$, with some halo-to-halo variation (not shown). For the low-mass `outlier' with $\Mtwohundred = 10^{10}\,\Msolar$, measured at $0.25 \times \Rtwohundred$, the radius where the outflow measurements exceed those in the \standard{} simulation occurs closer to the galaxy, at around $0.15 \times \Rtwohundred$. 

These measurements show that CRs sustain winds over larger distances, reflected in the higher mass loading factors at the virial radius in the \CRs{} simulations. In contrast, the \standard{} simulations display a steeper decline in mass loading with increasing distance. From theory, two limiting behaviors are expected for the relation between CR pressure and gas density. In the purely adiabatic case, CRs behave as a relativistic fluid with adiabatic index $\gamma_{\rm cr}=4/3$, which implies $P_{\rm cr} \propto \rho^{4/3}$. In the streaming-dominated case, flux-tube analyses show that CRs streaming at the Alfvén speed lead instead to $P_{\rm cr}\propto \rho^{2/3}$ (e.g. \citealt{Breitschwerdt1991}; see also Sect.~3.2.1 of the review by \citealt{Ruszkowski_Pfrommer2023}). These are idealised scalings derived in simplified setups where the transport physics can be modeled in a controlled way. In our cosmological simulations, where CR streaming is not treated explicitly but emulated through an Alfvén-cooling term, it is not obvious how the $P_{\rm cr}$--$\rho$ relation should appear. Nevertheless, the measured relation in Figure~\ref{fig:PcrRho} lies between these two limits at low densities, with the bulk of the gas in both the low-mass ($\Mtwohundred = 10^{10}\Msolar$) and high-mass ($\Mtwohundred = 10^{12}\Msolar$) halos tending toward the adiabatic expectation. In this low-density regime, the broader cosmological environment further contributes to setting the distribution of gas properties. At higher densities the situation is more complex for both halos, and the slope of the $P_{\rm cr}$--$\rho$ relation varies as processes such as radiative gas cooling, Coulomb and hadronic CR losses, magnetic field amplification, turbulent mixing, and the time variability of CR injection alter the scaling. In addition, CR energy is also dissipated over longer timescales because CR cooling is generally less efficient than the radiative cooling of a thermal plasma \citep{Ensslin+2007}. The combination of the softer adiabatic equation of state at low densities and the longer cooling timescales of CRs compared to a thermal plasma allows them to sustain winds as they propagate through the CGM, resulting in higher mass outflow rates, velocities, and mass loading at the virial radius in the \CRs{} simulations compared to the \standard{} simulations for lower-mass galaxies.

\textit{\textbf{Summary:}} In our simulations, CRs significantly impact the outflow properties of galaxies in lower mass halos. Near the galaxy ($0.25 \times \Rtwohundred$), the standard simulations show enhanced mass outflow rates, outflow velocities, and mass loading factors for galaxies with halo masses $\lesssim 10^{11.5}~\Msolar$ compared to those including CRs. However, at the virial radius, these properties are higher in the simulations with CRs, thus showing the opposite effect. This indicates that for our lower mass galaxies CRs help sustain winds over larger distances even though these galaxies exhibit lower SFRs in the \CRs{} simulations. For higher mass galaxies (halo masses above $10^{11.5}~\Msolar$), the effects of CRs on outflow properties are unchanged across different simulations, indicating that stochastic variations outweigh the impact of CRs.

This conclusion may be influenced by (1) our chosen CR transport model and (2) the use of the effective wind model in our simulations. On the one side, including CR streaming and diffusion within the self-confinement picture \citep{Thomas+Pfrommer2019,Thomas2023,Thomas+2024} could yield different CR transport speeds and momentum and energy transfer to the gas while CR scattering at externally driven turbulence \citep{Zweibel2017,Kempski2022,Ruszkowski_Pfrommer2023} could modify the CR feedback efficiency furthermore. On the other side, in the Auriga wind model, the local dark matter velocity dispersion, which generally increases with halo mass, determines the wind velocity. As a result, winds in lower mass galaxies tend to be slower, while those in higher mass galaxies are faster. This may result in CRs having little effect in higher mass galaxies because of the already fast effective winds. Models that incorporate more detailed feedback processes, rather than relying on effective prescriptions, could lead to different outcomes. Nevertheless, our results underscore the important role of CRs in shaping outflow properties and regulating star formation, particularly in lower mass systems.
%
\section{Variations in cosmic ray physics}
\label{sec:CRsVar}
In this Section we test the sensitivity of our results on CR parameter choices by running a representative sample of galaxies with variations in CR physics. This approach helps to assess how uncertainties within CR theory influence our findings. The variations we explore are:
\begin{enumerate}
    \item \textit{Acceleration Efficiency:} Our default CR simulation assumes a CR acceleration efficiency of $\zeta_{\text{SN}} = 0.1$, leading to a fixed amount of energy per unit of stellar mass formed (see Section~\ref{sec:surge}). We run 6 out of the 13 simulations across the whole mass range with a different acceleration efficiency of $\zeta_\mathrm{SN} = 0.05$. 
    \item \textit{Alfvén Cooling:} We run all the simulations with our fiducial CR model and without the Alfvén cooling term. In Equation~(\ref{eq:EnergyDensityEvol}), this corresponds to the term $|\boldsymbol{\varv}_\mathrm{A} \bcdot \bnabla P_\mathrm{CR} |$ (see Section~\ref{sec:surge}) so that these models correspond to pure CR diffusion on top of the advective CR transport with the gas.
    \item \textit{Diffusion Coefficient:} Our default CR simulations use a CR diffusion coefficient along the direction of the magnetic field of $\kappa = 10^{28} \, \text{cm}^2 \, \text{s}^{-1}$. We run four additional representative simulations across the mass range with $\kappa = 3 \times 10^{28} \, \text{cm}^2 \, \text{s}^{-1}$ and one simulation of a MW-like halo with $\kappa = 10^{29} \, \text{cm}^2 \, \text{s}^{-1}$.
\end{enumerate}
In summary, our analysis shows that changing the CR acceleration efficiency has only a minimal impact on the SFR, leading to a slightly higher SFR rate at lower acceleration efficiencies. This results in a marginally higher total stellar mass. Similarly, the effect on outflow properties is negligible. A likely explanation is that the system reaches a self-regulating state where the injection rate no longer significantly influences these properties.
\begin{figure*}
    \centering
	\includegraphics[width=0.7\linewidth]{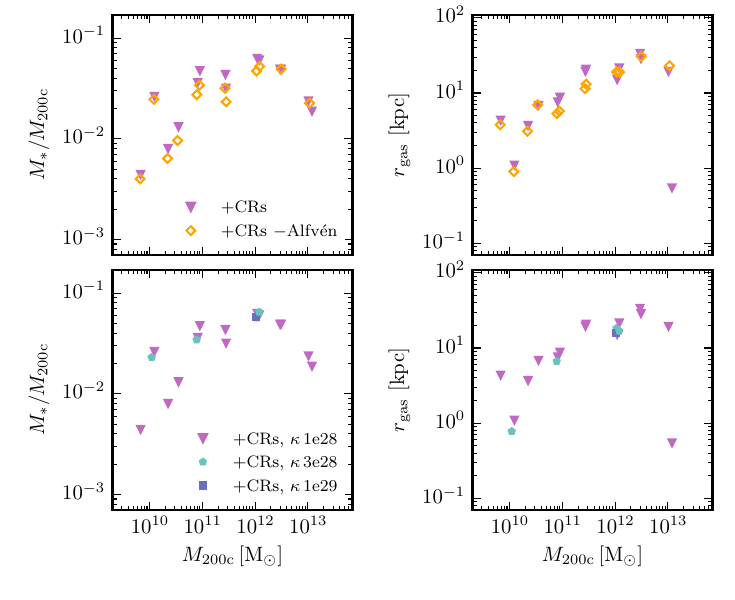}
    \caption{Stellar mass (left panels) and gas disc size (right panels) at $z = 0$ from all CR simulations as a function of halo mass. The top row shows the measurements for the standard \CRs{} and \noalfven{} simulations, while the bottom row displays those for the fiducial CR simulations as well as those with varying diffusion coefficients. Excluding Alfvén cooling or increasing the CR diffusion coefficient slightly reduces the stellar mass and gas disc size in galaxies with halo masses at or below that of MW-like systems.}
    \label{fig:VariationsDisk}
\end{figure*}

In contrast, we find that variations in the CR transport parameters -- specifically by changing the diffusion coefficient or including / excluding Alfvén cooling -- affects SFRs, CGM properties, and the dynamics of galactic outflows in galaxies with halo masses below $10^{12}~\Msolar$. We also find increased outflow velocities in halos of $10^{11} - 10^{12}~\Msolar$ when Alfvén cooling is turned off. These results highlight the strong sensitivity of low- and intermediate-mass galaxies to CR transport physics. Lower-mass galaxies, in particular, serve as valuable laboratories for studying these effects and refining CR transport models. This motivates further research into improving and constraining these models, which will ultimately enhance our understanding of the role CRs play in shaping galaxy evolution.  

The following subsections focus on the CR transport parameter variations, as these produce the most significant differences in galaxy and CGM properties.
\subsection{The effect of excluding Alfvén cooling} 
Alfvén cooling approximates the energy loss from CRs due to the damping of Alfvén waves excited by CR streaming. This process captures one of the key physical effects of CR streaming by partially converting CR energy into thermal energy, thus reducing the overall CR energy and pressure. However, the efficiency of this process remains uncertain and likely depends on local plasma conditions. Without the Alfvén cooling term, CRs in our simulations no longer lose energy through wave damping. In both cases, CRs can lose energy through other mechanisms such as adiabatic expansion, Coulomb cooling, and hadronic cooling (see Equation~\ref{eq:EnergyDensityEvol}). In this section, we examine the effect of excluding Alfvén cooling and focus on its influence on galaxy properties and gas dynamics.

Excluding Alfvén cooling leads to differences in the relative importance of CR pressure compared to thermal pressure within the galaxy as well as within its CGM. Figure~\ref{fig:Xcr} illustrates the ratio of CR to thermal pressure ($X_\mathrm{CR} = P_\mathrm{CR} / P_\mathrm{th}$) for four representative galaxies. The top panels show results from simulations that include Alfvén cooling (i.e., \CRs{}), while the bottom panels show simulations without it (i.e., \noalfven). 
\begin{figure*}
    \centering
	\includegraphics[width=\linewidth]{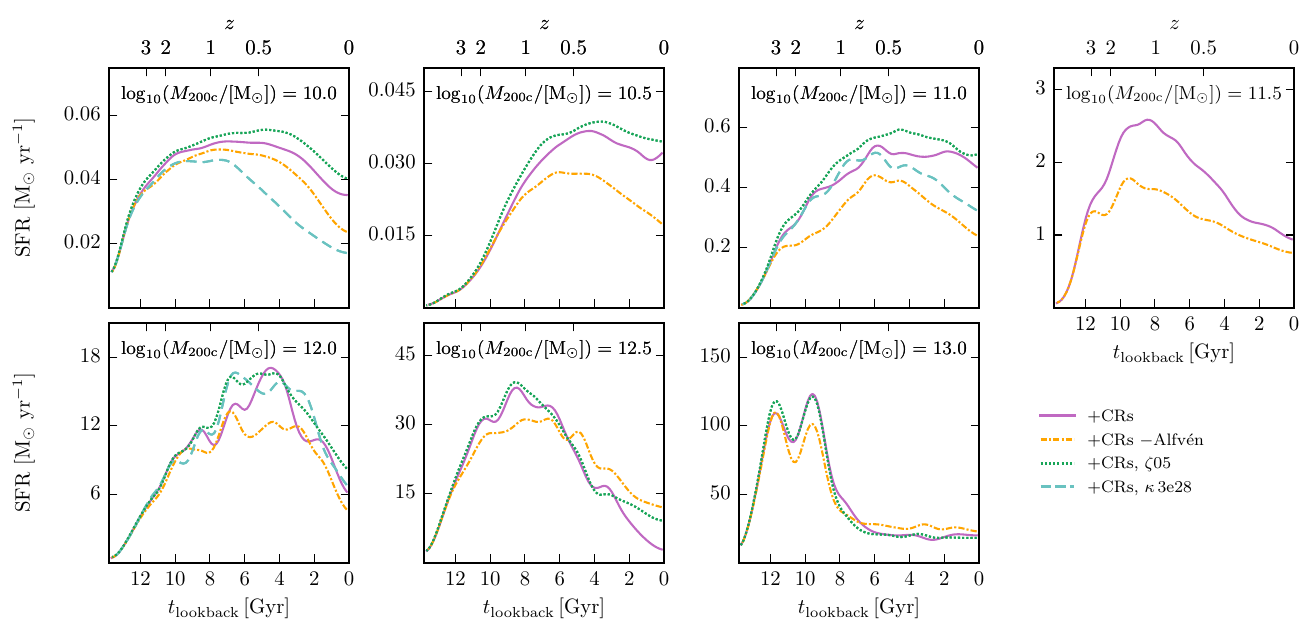}
    \caption{Comparison of SFHs for the different variations in CR physics: no Alfvén cooling (\noalfven), higher diffusion coefficient ($\kappa 
    =3\times 10^{28}\, \text{cm}^2\,\text{s}^{-1}$), lower CR acceleration efficiency ($\zeta_\mathrm{SN} = 0.05$). Changing the CR transport parameters alters the SFH of the lower mass galaxies (i.e., halo masses below $\Mtwohundred = 10^{12}~\Msolar$) with the most significant relative effect in the intermediate mass galaxies (i.e., halo masses between $\Mtwohundred = 10^{11}~\Msolar$ and $10^{11.5}~\Msolar$). The effect of Alfvén cooling on the SFR is more pronounced at higher redshift for the more massive galaxies of halo masses $\Mtwohundred = 10^{12}~\Msolar$ to $10^{12.5}~\Msolar$. At these redshifts, the halos were less massive, more closely resembling the present-day low-mass dwarfs in their response to CR physics, which likely made them more sensitive to its effects. Additionally, changes in the CGM environment at higher redshifts may have further contributed to this increased sensitivity. }
    \label{fig:VariationsSFH}
\end{figure*}

The relative importance of CR pressure compared to thermal pressure is visually different when including/excluding the Alfvén cooling term for all mass ranges probed. However, $\Xcr$ increases in different regions in and around the galaxy depending on the halo mass in these two different cases. For simplicity, we refer to the region within $0.25 \times \Rtwohundred$ as the \textit{galaxy and immediate vicinity} and the region beyond this radius as the \textit{extended CGM}. 

In the \noalfven{} simulations, CR pressure dominates over thermal pressure within the galaxy and its immediate vicinity for all halo masses. Notably, for halos with masses of $10^{12}~\Msolar$ or greater, CR pressure in the \noalfven{} simulations surpasses thermal pressure, reaching as high as $\Xcr \sim 40$ in the MW-like galaxy. This higher-than-unity $\Xcr$ region extends further for the MW-like galaxy compared to the higher mass galaxy. In the extended CGM of galaxies with halo masses equal to or greater than that of the MW, $\Xcr$ in the \noalfven{} simulations is only slightly higher than in the fiducial runs and remains well below unity.

For galaxies with halo mass of $10^{11}\,\Msolar$ and lower, CR pressure dominates over the thermal pressure within the galaxy and its vicinity for both the fiducial and \noalfven{} simulations. While $X_\mathrm{CR}$ is somewhat higher in the \noalfven{} runs, the difference remains small in this region. In the extended CGM, however, $X_\mathrm{CR}$ is much higher further away from the galaxy for $10^{11}\,\Msolar$ halos, and slightly elevated for the lowest-mass galaxy, when Alfvén cooling is excluded. 

This likely occurs because, without the Alfvén cooling term, CRs retain more of their energy as they are transported into the halo. As a result, they accumulate more effectively in the CGM and contribute significantly to the local pressure, especially in galaxies in the mass range of $10^{11}$--$10^{12}~\Msolar$. In these regions, hydrostatic equilibrium is maintained with a larger share of the support provided by CRs, so the required thermal pressure -- and thus the gas temperature -- is lower. Consequently, regions with increased $\Xcr$ exhibit much lower temperatures (not shown) -- most prominently in galaxies within the intermediate mass range. 

Excluding Alfvén cooling also affects star formation and galaxy structure, especially in lower-mass systems. Figure~\ref{fig:VariationsDisk} shows the stellar mass and gas disc size at $z=0$ as a function of halo mass for the different CR physics variations. In simulations without Alfvén cooling, galaxies with halo masses below $10^{12}~\Msolar$ have consistently lower stellar masses and develop slightly smaller gas discs compared to the fiducial CR simulations. 

For MW-like galaxies, the effect of Alfvén cooling on gas disc size is inconclusive, with some galaxies showing slightly smaller and others slightly larger discs (see Figure~\ref{fig:VariationsDisk}). Note that the galaxy analyzed by \citet{Buck2020} has a smaller gas disc size in the simulations without Alfvén cooling. They argue that the exclusion of Alfvén cooling changes the outflow geometry and suppresses the accretion of high angular momentum gas, particularly at late times, leading to smaller gas discs. This explanation is plausible, but we will need to test it in future work and understand where the differences between individual galaxies arise.

Figure~\ref{fig:VariationsSFH} shows the SFHs for the same set of simulations, highlighting that the reduction in SFR is most pronounced for halos between $10^{10.5}~\Msolar$ and $10^{11.5}~\Msolar$. Consequently, the differences in stellar mass at $z=0$ are also most prominent in this mass range. For MW-like galaxies, the effect of Alfvén cooling on the SFR is more pronounced at higher redshift. This is potentially because at higher redshift these MW-like galaxies were smaller and more similar to lower-mass dwarfs. Over time, as these galaxies gain mass, the difference in SFR between runs with and without Alfvén cooling becomes less pronounced, possibly because the CR pressure builds up to similar levels in both cases. A similar trend emerges for more massive galaxies (i.e., $\Mtwohundred = 10^{13}\,\Msolar$), although less pronounced. We will examine the impact of CRs on the galaxies and the CGM throughout cosmic time in more depth in a follow-up paper. 
\begin{figure*}
    \centering
	\includegraphics[width=\linewidth]{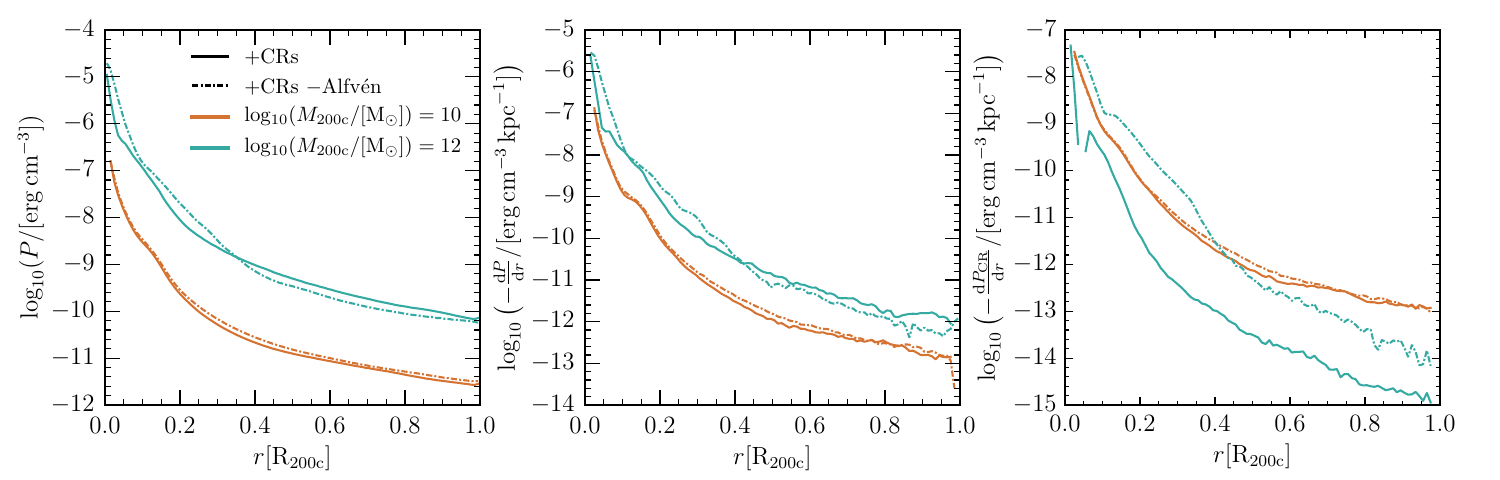}
    \caption{Radial profiles of total pressure (CR + thermal + magnetic) and pressure gradients for halos of mass $10^{10}~\Msolar$ (ochre) and $10^{12}~\Msolar$ (mint), comparing simulations with CRs (\CRs{}, solid lines) and with CRs but without Alfvén cooling (\noalfven{}, dash-dotted lines). The three columns show, from left to right: total pressure $P$, total pressure gradient $\log_{10}(-\rmn{d}P/\rmn{d}r)$, and CR pressure gradient $\log_{10}(-\rmn{d}P_{\mathrm{CR}}/\rmn{d}r)$. All profiles are spherically averaged and computed excluding satellite galaxies and star-forming ISM gas. The profiles are averaged over the last $\sim$1~Gyr of evolution.}
    \label{fig:PressureGrad}
\end{figure*}

The reduced SFRs in the \noalfven{} simulations could result from two contributing effects: (i) more efficient outward transport of gas, and (ii) suppression of inflows onto the disc due to increased pressure support in the halo. As shown in Figure~\ref{fig:PressureGrad}, the total pressure -- defined here as the sum of CR, thermal, and magnetic field pressures -- in the \noalfven{} simulations is noticeably higher in the inner regions compared to the fiducial runs for the $10^{12}\,\Msolar$ halo. However, the profiles converge at $0.25\,\Rtwohundred$, and the differences beyond this radius are smaller. This suggests that the enhanced total pressure primarily affects regions close to the galaxy and may reduce inflow rates in the inner halo. 

\begin{figure*}
    \centering
	\includegraphics[width=\linewidth]{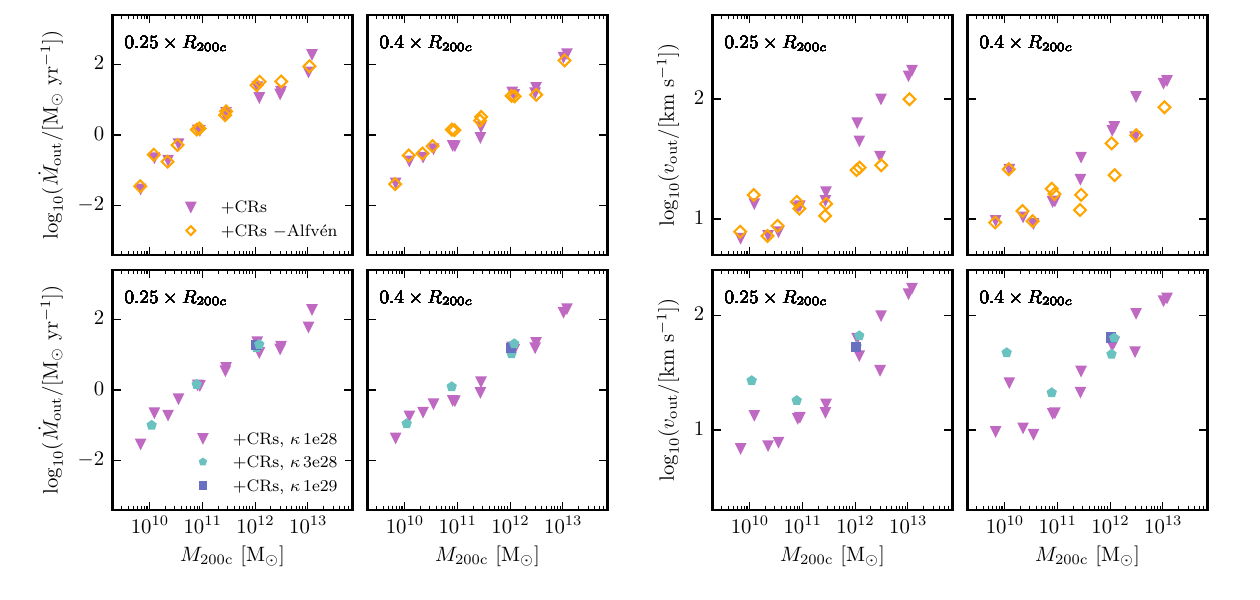}
    \caption{Outflow properties for different physics variations at two different radii as a function of halo mass for all cosmological simulations and at $z = 0$. The top row shows the measurements from the \CRs{} and \noalfven{} simulations, while the bottom row displays the measurements from the \CRs{} simulations and those with varying diffusion coefficients. On the left/right hand side are two panels showing the mass outflow rate/outflow velocity for the different simulations measured at $0.25 \times \Rtwohundred$ and $0.4 \times \Rtwohundred$, respectively. Both the mass outflow rates and outflow velocities are averaged over 10 snapshots, covering approximately 1~Gyr or the redshift range $z=0.1 - 0$. Excluding Alfvén cooling or changing the CR diffusion term affects outflow properties across the different halo masses probed. When Alfvén cooling is excluded, the mass outflow rates are notably altered further away from the galactic disc, while outflow velocities are affected for all halo masses and both close to and further away from the disc for halo masses below $\Mtwohundred = 10^{12}~\Msolar$. Changes in the diffusion coefficient primarily impact mass outflow rates and outflow velocities for galaxies with halo masses lower than or equal to $\Mtwohundred = 10^{11}~\Msolar$.}
    \label{fig:VariationsOutflow}
\end{figure*}

Figure~\ref{fig:VariationsOutflow} presents the mass outflow rates (two leftmost panels) and outflow velocities (two rightmost panels) for the \standard{} and \noalfven{} simulations, measured at two radii: $0.25 \times \Rtwohundred$ (top row) and $0.4 \times \Rtwohundred$ (bottom row).

The larger CR pressure gradient in the \noalfven{} simulation (see Figure~\ref{fig:PressureGrad}) for the more massive galaxies allows CRs to exert a stronger force on the surrounding gas, resulting in a greater push on the material. This drives more gas out of the galaxy, leading to a higher mass outflow rate in the vicinity of the more massive galaxies (i.e., halo mass $10^{12} - 10^{13}~\Msolar$). However, the mass loading factor (not shown) in the \noalfven{} simulation is lower near the galaxy compared to the \CRs{} runs, while increasing in the extended CGM. This variation suggests that, without Alfvén cooling, more gas is transported to the outer regions of the CGM rather than retained near the galaxy. At the same time, the mass inflow rate to the galaxy (not shown) is also decreased, likely due to the increased total pressure support in the inner halo, reducing the amount of gas available for star formation (see Figure~\ref{fig:VariationsSFH}). In the \noalfven{} runs, the outflowing gas moves at lower velocities than in the fiducial runs with Alfvén cooling, but because a larger amount of gas is accelerated, the total mass transported into the CGM is higher. This interpretation is supported by the increased gas density near the disc and the higher mass loading factors in the outer halo (both not shown). Further from the galaxy, the $\Xcr$ values in the two simulations converge, leading to similar outflow velocities and mass outflow rates at large radii with or without Alfvén cooling.

For intermediate-mass galaxies (i.e., halo masses below $\Mtwohundred = 10^{11} - 10^{11.5}~\Msolar$), the mass outflow rate in the \noalfven{} simulations is only increased around $0.4 \times \Rtwohundred$. This increase corresponds to higher $\Xcr$ values and a slightly elevated CR pressure gradient (see Figure~\ref{fig:PressureGrad}) compared to the fiducial CR simulations. At larger radii ($\sim\Rtwohundred$), the differences in CR energy losses due to Alfvén cooling become less significant. Consistent with this, the $\Xcr$ values, mass outflow rates, and outflow velocities converge at the virial radius (not shown), regardless of whether Alfvén cooling is included.
\subsection{The effect of changing the diffusion coefficient}
We next examine the impact of varying the CR diffusion coefficient on galaxy properties across the simulated halo mass range. Increasing the diffusion coefficient reduces the SFR in lower-mass galaxies, leading to a lower final stellar mass (see Figures~\ref{fig:VariationsSFH} and~\ref{fig:VariationsDisk}). For MW-like galaxies, however, the increased diffusion coefficient leads to variations in the star formation history that are consistent with stochastic fluctuations rather than a systematic trend. The resulting total stellar mass is, in addition, comparable for the MW mass galaxies. 

The diffusion coefficient also affects the gas disc sizes, particularly at smaller halo masses, likely through its influence on outflow properties (see Figure~\ref{fig:VariationsDisk}). In addition, higher diffusion coefficients increase the velocity of the outflowing gas for galaxies with halo masses less than or equal to $\Mtwohundred =10^{11}~\Msolar$, both near the galaxy and in the extended CGM (see Figure~\ref{fig:VariationsOutflow}). However, for the lowest-mass halo in our sample, mass outflow rates are slightly decreased in the run with a higher diffusion coefficient, both near and further away from the galaxy. In contrast, for the galaxy in a halo of $\Mtwohundred = 10^{11}~\Msolar$, the mass outflow rate near the disc is unchanged between diffusion coefficients, but increases slightly at larger radii in the run with the higher diffusion coefficient. Note that these results are based on a single galaxy per mass bin and may not be representative of all low-mass systems.

In these lower-mass galaxies, mass loading factors (not shown) decrease with increasing diffusion speed, both near the disc and in the extended CGM. This trend is consistent with previous findings in the literature \citep[see discussion on page 74 in the review by][]{Ruszkowski_Pfrommer2023}, where higher diffusion coefficients are associated with faster outflows and reduced mass loadings. This behaviour aligns with expectations, because faster diffusion reduces CR pressures and pressure gradients, weakening the outflows, and lowering the mass loading.

Interestingly, we observe no significant change in outflow properties for galaxies with MW mass ($\sim10^{12}~\Msolar$), when varying the diffusion coefficient from $10^{28}\, \text{cm}^2 \, \text{s}^{-1}$ and $ 10^{29}\, \text{cm}^2 \, \text{s}^{-1}$ to $3\times 10^{28}\, \text{cm}^2 \, \text{s}^{-1}$ and $ 10^{29}\, \text{cm}^2 \, \text{s}^{-1}$, respectively. This may arise from several factors. First, the deeper gravitational wells of MW-like and higher-mass galaxies could reduce their sensitivity to changes in the diffusion coefficient. Second, the faster winds generated by the Auriga wind model at this mass scale may make these galaxies less sensitive to changes in CR diffusion. Third, Alfvén cooling may be strong enough to reduce the impact of CRs regardless of the diffusion coefficient. Finally, the higher thermal pressure in more massive haloes makes it harder for CRs to dominate the pressure budget, and the increasing importance of AGN feedback in these systems could further reduces the effects of CR diffusion on outflows.

\section{Discussion}
\label{sec:discussion}
Our simulations demonstrate that CRs can influence galaxy properties, CGM structure, star formation, and outflows, with their effects being most pronounced in lower-mass galaxies. However, differences in feedback models, ISM treatments, and CR transport prescriptions across simulations can lead to contrasting conclusions about both the strength of CR-driven feedback and the dominant physical mechanisms by which CRs affect gas dynamics. 

First, we compare our findings to previous studies, highlighting agreements and discrepancies in how CRs regulate galaxy properties, the CGM, star formation, and outflows. Then, we discuss key model uncertainties, including ISM and feedback models and the CR transport prescription, particularly compared to two-moment models.

\subsection{Comparison to Previous Work} 

In this Section, we compare our findings to prior work, focusing on four key areas discussed in this paper: (1) global galaxy properties, (2) the effect of CRs on gas properties, (3) CR-driven regulation of star formation, and (4) the impact of CRs on outflow dynamics.

\subsubsection{Global Galaxy Properties at the Present Time}
Our results show that CRs affect both the morphology and stellar content of galaxies across the full halo mass range, with the strongest overall impact in lower-mass systems. In halos below $10^{12}~\Msolar$, CRs reduce the stellar mass by up to a factor of two and also alter the morphology. In more massive galaxies, CRs primarily affect morphology without significantly changing the total stellar mass. Across all masses, CRs reduce the gas disc size and the stellar half-light radius, while the stellar half-mass radius remains largely unchanged.

Previous studies have also examined how CRs influence galaxy morphology and stellar mass growth. \citet{Hopkins+2021b} found in cosmological zoom simulations that CRs impact galaxy morphology, but that the strength of this effects depends on the CR transport model. They reported that, across dwarf to Milky Way-mass galaxies, CRs suppress stellar mass growth by factors of 2–3, and that the strongest effects occur in MW-like galaxies, where CR pressure significantly alters gas distributions and CGM structure. Our results are broadly consistent in magnitude but differ in the mass range where CRs are most effective: we find that CRs have the strongest impact in lower-mass galaxies, where they suppress stellar mass growth by up to a factor of two and also modify morphological properties.

One key difference between our simulations and those of \citet{Hopkins+2021b} is the choice of the CR diffusion coefficient. Our work adopts a value of $\kappa = 1 \times 10^{28}$ cm$^2$ s$^{-1}$ in our fiducial run, while \citet{Hopkins2019} use a significantly higher value of $\kappa = 3 \times 10^{29}$ cm$^2$ s$^{-1}$ in their fiducial model. We refer to this as their fiducial model because it provides the best match between gamma-ray emission from isolated dwarfs and observations of satellite galaxies \citep{Hopkins2019, Chan+2019}. The need for a higher diffusion coefficient in their simulations in order to match observations likely also reflects the differences in magnetic field strength: at $z=0$, the magnetic fields in the simulations by \citet{Hopkins2019} are significantly weaker than in our simulations (by a factor of 10 in field strength below our results). With such low field strengths, the Alfvén speed is small and the associated wave damping rates are negligible, so CRs lose little energy through this channel regardless of the implemented physics. In addition to differences in the diffusion coefficient and magnetic field strengths, the simulations by \citet{Hopkins2019} and our simulations also differ in the treatment of the ISM and the feedback model. For a detailed discussion of how differences in the ISM models between the FIRE-2 and Auriga simulations influence the interpretation and effectiveness of a given diffusion coefficient, see \citet{Buck2020}.

Other studies have also investigated how CRs regulate disk structure. Using cosmological zoom-in simulations, \citet{Salem+2014b} found that simulations with CRs produce thinner and more extended stellar disks compared to purely thermal feedback models. Their simulations used isotropic diffusion with a constant diffusion coefficient and omit CR energy losses (such as Coulomb or hadronic cooling, and CR losses due to excitation of Alfvén waves). In contrast, our simulations include anisotropic diffusion and account for Alfvén cooling, allowing for more realistic CR pressure gradients and their impact on the gas. As our results show, variations in CR transport properties affect the morphology of galactic discs, as well as star formation, outflow properties, and CGM structure. 

In contrast, \citet{Farcy+2022} studied isolated galaxies embedded in dark matter haloes of $10^{10}, 10^{11}, 10^{12}~\Msolar$, and found that CRs suppress star formation while leading to thicker and more stable galactic disks with a smoother gas distribution. In our simulations, we do not visually find systematically thicker gas discs, but we do see a smoother gas distribution in the presence of CRs (see Figure~\ref{fig:gasprojection}). Their simulations also included stellar radiation feedback, which could modify CR-driven pressure effects and contribute to disk thickening. Unlike \citet{Farcy+2022}, we use an effective wind model in which winds are launched independently of the local gas properties and CR pressure. CRs only begin interacting with the outflowing material after the winds recouple, which may alter the way CRs affect the CGM and disc structure. However, as discussed in \citet{Buck2020}, CRs also modify gas flows in the CGM, which subsequently alters the angular momentum distribution of accreted gas and can lead to smaller gaseous discs. The interplay between CRs, CGM gas flows, and angular momentum distribution in our simulations provides an additional mechanism for shaping disk morphology, beyond the direct impact of CRs on the internal gas structure of the disc. This process is only modelled self-consistently in full cosmological simulations, but not captured in the idealised galaxy simulations of \citet{Farcy+2022}. 

In summary, differences in disk structure and stellar mass growth across simulations stem from variations in feedback prescriptions, ISM treatment, and CR transport modeling. While all of the studies discussed above find that CRs suppress stellar mass growth, the magnitude of this effect and their halo mass dependence varies between different studies. In addition, it is important to consider the larger-scale context in which galaxies evolve, as CRs influence not only processes within the galaxy itself but also CGM gas flows. For instance, \citet{Buck2020} showed that CR-driven modifications to CGM dynamics can alter the angular momentum of accreted gas and lead to more compact gaseous discs. This highlights the need for cosmological simulations to fully capture CR-driven effects on disc morphology and stellar mass regulation. 

\subsubsection{Influence of CRs on Gas Properties}
CRs influence the CGM by altering its temperature, density, metallicity, and magnetic field structure. Several previous studies have examined these effects, highlighting the role of CRs in regulating CGM properties. One of the earliest investigations into these effects was conducted by \citet{Salem+2016}, who studied a $10^{12}~\Msolar$ halo and found that CR-driven outflows efficiently transport metals and magnetic fields into the CGM. Their results suggested that CR pressure provides additional support for cool, low-density gas, reducing the reliance on thermal pressure for pressure equilibrium.

Earlier work using cosmological zoom-in simulations also highlighted the importance of CRs in shaping the CGM. \citet{Ji+2020} examined CR-dominated halos in their simulations, showing that CR pressure can exceed thermal pressure in MW-mass galaxies at low redshift, producing a cool, volume-filling CGM. Specifically, their simulations show a reduction in CGM temperatures from above $10^6$~K to around $10^5$~K. In addition, \citet{Hopkins2019} showed, using the same simulations, that the strength of CR pressure support is highly sensitive to the assumed diffusion coefficient, with lower values leading to stronger CR effects. In their simulations, \citet{Ji+2020} adopt a fiducial value of $\kappa = 3 \times 10^{29}\,\mathrm{cm}^2\,\mathrm{s}^{-1}$, calibrated to match CR population models in the Milky Way and gamma-ray observations of nearby galaxies \citep{Hopkins+2021a, Chan+2019}. In contrast, our fiducial simulations do not produce such a drastic reduction in CGM temperature in this mass range, resulting in a systematically hotter CGM than reported in the fiducial simulations by \citet{Hopkins+2021b} and \citet{Ji+2020}.

One possible reason for these discrepancies could be differences in magnetic field strengths and CR diffusion coefficients between the simulations by \citet{Hopkins+2021a} and our own. Their simulations yield weaker magnetic fields, particularly in the CGM \citep[see][]{Hopkins+2021a}, which results in lower Alfvén cooling and thus reduces CR energy losses as they are transported to large radii. In contrast, our stronger magnetic fields and lower CR diffusion coefficient create a different balance between CR energy loss and gas cooling. This leads to a hotter CGM compared to the simulations studied in \citet{Hopkins+2021b} and \citet{Ji+2020}, and is more consistent with X-ray observations of the MW hot halo \citep[e.g.,][]{Fang+2013, Faerman+2017}. These differences underscore the fact that CGM properties are sensitive to magnetic field strength and the treatment of CR transport -- in particular to the inclusion of Alfvén cooling -- highlighting the importance of accurately capturing these processes in cosmological galaxy simulations. 

More recently, \citet{Farcy+2022} analysed isolated galaxy simulations spanning halo masses from $10^{10}~\Msolar$ to $10^{12}~\Msolar$ and found that CRs significantly influence the density and temperature structure of the CGM. Their results emphasised the importance of the choice of CR diffusion coefficients, demonstrating that higher diffusion coefficients smooth CR pressure gradients, thereby modifying the thermal and density structure of the CGM. Although our study does not explicitly explore how the CGM properties vary in response to changes in the diffusion coefficient, preliminary analysis shows no strong visual differences. However, further quantitative analysis would be required to thoroughly compare our results with those of \citet{Farcy+2022}. Additionally, our cosmological simulations naturally incorporate environmental effects and gas accretion, all of which likely affect the impact of CRs and potentially lead to differences in how CR transport shapes CGM properties compared to the findings of \citet{Farcy+2022}.

Additionally, our simulations show that CR-driven outflows efficiently transport metals and magnetic fields into the CGM, particularly in galaxies with halo masses below $10^{12}\,\Msolar$. \citet{Salem+2016} found a similar effect in a more massive system with a halo mass of $10^{12}\,\Msolar$, but their study did not explore the lower-mass regime, where we find the strongest impact. Our results also align, by construction, with the CRdiffAlfvén model presented by \citet{Buck2020}, which corresponds to our fiducial CR run and uses the same Auriga galaxy formation model. They showed that Alfvén cooling significantly raises CGM temperatures compared to models without Alfvén losses (see also Section~\ref{sec:CRsVar} for a discussion of a similar result in our simulations). This highlights an important difference from the simulations by \citet{Salem+2016}, which did not include Alfvén cooling and thus could not capture the associated CR energy-loss process. In our simulations, this process can modify CR pressure support in the CGM and, in models without Alfvén cooling, lead to lower CGM temperatures and altered gas dynamics.

In summary, differences in CGM properties between simulation suites can be traced to variations in magnetic field strengths, CR diffusion coefficients, the inclusion of Alfvén wave losses, as well as differences in feedback prescriptions. A key conclusion of our work is that accurately capturing the role of CRs in shaping the CGM requires detailed CR transport physics, including multiple transport mechanisms such as anisotropic diffusion and Alfvén wave cooling as is realized in two-moment descriptions of CR transport \citep{Jiang+Oh2018,Thomas+Pfrommer2019}. Also important is the cosmological context, which self-consistently models gas accretion and environmental effects that influence how CRs interact with the CGM. As CRs are sensitive to a wide range of physical conditions, they offer a promising diagnostic of otherwise difficult-to-observe CGM properties, for example through indirect tracers such as gamma-ray emission from hadronic interactions or synchrotron radiation from CR electrons.

\subsubsection{Effect of Cosmic Rays on Star Formation}
Our simulations demonstrate that CRs significantly reduce star formation in galaxies residing in halos with masses below $10^{12}\,\Msolar$. The suppression is strongest in lower mass halos, with star formation rates reduced by a factor of two compared to simulations without CRs. In contrast, for halos more massive than $10^{12}\,\Msolar$, CRs have minimal impact on star formation. 

Earlier studies have consistently found that CRs reduce star formation, although the magnitude and mass dependence of this suppression vary between simulations. \citet{Salem+Bryan2014} showed that CRs can effectively reduce star formation by redistributing gas away from dense, star-forming regions through additional pressure support, leading to lower overall star formation efficiency. Similarly, \citet{Farcy+2022} found significant CR-driven star formation suppression in isolated dwarf galaxies, with diminishing effectiveness at higher galaxy/halo masses -- in line with our findings. They attributed this effect to CR pressure smoothing out gas density in regions where stars form and explode, thereby reducing the collapse of gas into stars.

On the other hand, the FIRE-2 simulations \citep[e.g.][]{Hopkins+2021b} reported a reduction in star formation due to CRs across all halo masses and highlighted that the impact of CRs depends strongly on the adopted CR diffusion model. In their MW-like galaxies, \citet{Hopkins2019} find that low diffusivity ($\kappa = 3 \times 10^{28}\,\mathrm{cm}^2\,\mathrm{s}^{-1}$) traps CRs in dense gas, where they lose energy to hadronic interactions and have little dynamical effect, while high diffusivity ($\kappa > 3 \times 10^{29}\,\mathrm{cm}^2\,\mathrm{s}^{-1}$) lets CRs escape too rapidly to build up halo pressure. Their fiducial model adopts $\kappa = 3 \times 10^{29}\,\mathrm{cm}^2\,\mathrm{s}^{-1}$, which they argue strikes a balance between these two regimes and provides the best match to gamma-ray constraints. Note that in \citet{Hopkins2019}, $\kappa = 3 \times 10^{28}\,\mathrm{cm}^2\,\mathrm{s}^{-1}$ is described as the low-diffusivity regime, whereas in our simulations this value corresponds to the higher-diffusion test case. Despite this overlap in value, the physical context differs -- our simulations have stronger magnetic fields and a different ISM/feedback model -- so CRs do not remain as trapped in dense gas. For MW-mass galaxies, both their low-diffusion runs and our $\kappa = 3 \times 10^{28}\,\mathrm{cm}^2\,\mathrm{s}^{-1}$ simulations show little suppression of star formation, suggesting similar trends in SFR at this mass scale despite differences in other galaxy and CGM properties. In a multi-phase ISM, strong ion-neutral damping of Alfvén waves should couple CRs weaker to the surrounding plasma and weaken CR feedback at densities $\gtrsim 10^{-2} \mathrm{cm}^{-3}$ \citep{Armillotta+2022,Armillotta+2024,Thomas+2024,Sike+2025}. This would suggest a revision of the interpretation of the FIRE-2 results on how CR feedback efficiency depends on the diffusion coefficient.

The picture differs for lower-mass galaxies. Our fiducial simulations adopt a lower diffusion coefficient ($\kappa = 1 \times 10^{28}\,\mathrm{cm}^2\,\mathrm{s}^{-1}$), which still falls in the ``low-diffusivity'' regime defined by \citet{Hopkins2019}, yet produces a clear suppression of star formation in galaxies below $10^{12}~\Msolar$. Increasing the diffusion coefficient to $\kappa = 3 \times 10^{28}\,\mathrm{cm}^2\,\mathrm{s}^{-1}$ strengthens this suppression in our simulations for this halo mass scale. In contrast, \citet{Hopkins2019} find little suppression in their low-diffusion runs for dwarf galaxies. 

The varying results suggest a complex, mass-dependent interplay between diffusion strength and star formation suppression due to CRs and indicates that other factors, such as the ISM model, feedback implementation, and magnetic field strength also contribute to the observed differences. 

\subsubsection{Dynamical Effect of CRs on Outflow Properties}
In our simulations, CRs influence galactic outflows and help sustain winds over larger distances in galaxies with halo masses below that of the Milky Way. This is quantitatively consistent with previous studies, although some of them also report significant CR effects in MW-like systems \citep[e.g.,][]{Hopkins+2021a}. 

For instance, cosmological simulations by \citet{Salem+2014b} and isolated galaxy disc simulations by \citet{Salem+Bryan2014} both demonstrated that CR-driven outflows produce strong multiphase winds with mass-loading factors around unity, highlighting the importance of CR pressure gradients and diffusion in effectively launching and smoothly accelerating these outflows. 

\citet{Farcy+2022}, using isolated galaxy simulations, found that CRs substantially increase mass-loading factors, particularly in lower-mass galaxies, and drive outflows that are both colder and have higher mass outflow rates compared to those produced by purely thermal feedback. Their study emphasized the sensitivity of outflow properties to the CR diffusion coefficient, showing that a higher diffusion coefficient enhances total mass outflow rates and shifts outflow temperature distributions toward hotter gas. In contrast, lower diffusion coefficients lead to stronger confinement of CRs within galaxies, resulting in higher mass-loading factors but slower, colder outflows. In our cosmological simulations, we find a similar dependence on the CR diffusion coefficient, where a higher coefficient leads to faster but lower-mass loaded outflows, particularly in lower-mass halos. 

In contrast, the FIRE-2 simulations \citep[i.e.,][]{Hopkins+2021a} found that CR-driven outflows in MW-like galaxies become coherent, cool, and volume-filling, extending to large scales. In their simulations, these effects are smaller at lower halo masses or higher redshifts because of changing CR-to-thermal pressure ratios. Their studies also highlighted the sensitivity of outflow characteristics to CR transport parameters, particularly the diffusion coefficient. In their fiducial model, they adopted a substantially higher diffusion coefficient compared to our fiducial simulations. In addition, their simulations produced much weaker magnetic fields than ours. These differences influence both the CGM properties and the nature of the outflows. \citet{Ji+2020} find, for MW-like galaxies, a much smoother, lower-temperature CGM, which facilitates the development of slow, volume-filling outflows. In contrast, our fiducial simulations maintain, again for MW-like galaxies, a more structured CGM with higher fraction of hot gas. CR-driven winds are, however, primarily present in lower-mass galaxies, where they show greater variability in their structures (see Figure~\ref{fig:velslice}).

We note that our fiducial model differs from the fiducial model adopted in \citet{Hopkins2019, Hopkins+2021a}. Our results further demonstrate that CR-driven outflows are highly sensitive to variations in CR transport physics. For instance, increasing the CR diffusion coefficient leads to higher wind velocities and lower mass-loading factors in low-mass galaxies, both near the disc and in the extended CGM, as CR energy spreads more efficiently throughout the system. Additionally, turning off Alfvén cooling enhances the mass-loading factor in our lower-mass halos, reinforcing the role of CR energy losses in shaping outflows. Interestingly, the CGM properties in the fiducial simulations by \citet{Ji+2020} best resemble those of our \noalfven{} runs. The lower magnetic field strength in their simulations causes Alfvén cooling to be strongly reduced compared to ours, making their runs more similar to our \noalfven{} simulations. 

Overall, both the FIRE-2-based and our Auriga-based simulation frameworks find that CRs contribute to sustaining winds, but the specific nature of these winds depends on the underlying physical models. The simulations by \citet{Hopkins2019, Hopkins+2021a} predict lower CGM temperatures than estimates from X-ray observations of the Milky Way halo \citep[e.g.,][]{Fang+2013, Faerman+2017}, whereas our simulations maintain a hotter CGM. These differences emphasise that the interplay between CR transport physics, magnetic field strengths, and CGM properties plays a crucial role in shaping galactic outflows.

\subsection{Model Uncertainties}
While our simulations provide valuable insights into the role of CRs in the evolution of galaxies, several modeling choices introduce uncertainties that may impact our conclusions. Key sources of uncertainty include the treatment of the ISM in our simulations, the effective feedback model, and the CR transport model. In this section, we discuss these aspects and, in addition, also compare our results to findings from two-moment CR transport models.

\subsubsection{ISM Model and its Influence on our Results}
Our simulations employ an effective model for the ISM \citep{SFR_paper}, also used in the Auriga simulations. It results in a relatively smooth gas distribution in the ISM, without high-density, low-temperature peaks typically seen in fully multiphase ISM models \citep[see e.g.,][]{Marinacci2019, Bieri2023}. Although this approach is well calibrated to reproduce the observed Kennicutt-Schmidt relation, we use it unchanged when adding CRs to focus on differential effects. The addition of MHD has been shown not to significantly alter the Kennicutt–Schmidt relation \citep[see Fig.~A1 in][]{Whittingham2021}, and we likewise do not expect substantial changes with CRs.   

The nature of the ISM model influences several aspects of our results. A fully multiphase ISM may confine CRs more strongly within dense star-forming regions, creating steeper local pressure gradients that could enhance the impact of the CRs on both star formation and outflows \citep[see also][]{Farcy+2022}. However, this effect can be counteracted by ion–neutral damping, which reduces CR confinement in partially neutral gas \citep{Sike+2025}. Since our model does not explicitly resolve cold and dense gas clouds, the interaction between CRs and the dense ISM phase may differ compared to simulations with a fully resolved multiphase structure, potentially altering CR escape into the CGM and the resulting impact on star formation. This likely also affects outflows: a smoother ISM allows CRs to couple more uniformly to the gas and drive more coherent winds, while a multiphase ISM with dense clouds may produce more irregular wind structures due to local variations in CR confinement and pressure gradients.

Recent work by \citet{Thomas+2024} employs a multiphase ISM model that explicitly resolves cold and dense clouds, providing a more detailed treatment of CR interactions with the dense gas phase. Their study also adopted a two-moment CR transport scheme, which models CR streaming along magnetic field lines and accounts for energy exchange with self-excited Alfvén waves in the local gas environment. They find that the majority of the CR energy injected by supernova remnants escapes the ISM and builds up in a thin region around the disc–halo interface, where it dominates the local pressure and contributes to driving galactic outflows. Despite differences in ISM modeling and CR transport prescriptions, the overall CR behavior in our simulations -- efficient escape into the CGM and contribution to outflows -- is broadly consistent with their findings. However, while \citet{Thomas+2024} report suppression of star formation and enhanced mass loading in MW-like galaxies, we do not find strong effects at MW masses, with CRs having a much larger impact in lower-mass systems. The fact that CRs efficiently escape into the CGM in both models suggests that large-scale CR dynamics are relatively robust to differences in ISM treatment. However, further investigation is needed to assess how incorporating a fully multiphase ISM in a cosmological context would influence the detailed coupling between CRs and outflowing gas.

\subsubsection{Feedback Model}
The feedback model in our simulations follows the Auriga framework, originally calibrated to reproduce the realistic properties of Milky Way-mass galaxies without CRs. In this model, stellar feedback launches galactic winds, with the initial wind velocities scaling with the local dark matter velocity dispersion. This results in lower wind launch velocities in low-mass halos, with higher launch velocities in MW-mass and more massive systems; although the winds become less efficient again at removing gas in the most massive halos. Additionally, in higher-mass galaxies, AGN feedback contributes to regulating star formation and outflows.

In our simulations, CRs do not have a systematic impact on stellar masses, star formation, or outflow properties in MW-like and higher-mass galaxies, with differences consistent with stochastic variations. Several factors may contribute to this. First, the deeper gravitational potential wells of these galaxies could make them less sensitive to changes in CR pressure. Second, in MW-mass halos, the relatively strong feedback-driven winds already present in the Auriga model may mask any additional effects from CRs. In contrast, in the most massive galaxies, AGN feedback dominates the regulation of gas inflows and outflows, again potentially overshadowing CR-driven processes. At the same time, because CRs do not significantly enhance winds in massive halos, this could suggest that even if both CRs and stellar feedback were modeled self-consistently, the resulting winds would not be substantially stronger than the feedback-driven winds already present in the Auriga model.

In lower-mass galaxies, CRs have a stronger influence on outflows and CGM structure. This suggests that the ability of CRs to drive outflows and regulate gas flows is mass-dependent, with CRs possibly playing a more significant role where stellar feedback alone is slower. However, since the Auriga feedback model was calibrated without CRs at MW-mass scales, some of the regulatory effects that CRs provide in other simulations may already be incorporated into the model's existing stellar and AGN-driven outflows. This could reduce the additional impact of CRs on star formation and galactic winds in our simulations.

\subsubsection{The CR-transport Model}
The impact of CRs on galaxy evolution also depends on how CR transport is modeled. Our simulations use a gray single-moment approach that evolves the total CR energy as a fluid-like component. CRs are advected with the gas and diffuse according to a spatially constant diffusion coefficient. We also include Alfvén cooling to emulate the effects of CR streaming. While this model captures key aspects of CR feedback -- such as their ability to transport energy and sustain galactic winds over larger distances in lower-mass galaxies -- it does not evolve the CR flux (i.e, the second moment of the CR distribution) self-consistently. This limits the accuracy of CR transport in multiphase environments, where local flux directions and gradients can vary significantly. 

Recent studies have explored more advanced two-moment CR transport models, which evolve both CR energy and flux, enabling more accurate modeling of CR streaming, diffusion, and their interactions with magnetic fields.
\citet{Thomas+2023, Thomas+2024} focused on isolated MW-like galaxies and found that CR pressure gradients have little impact on the star-forming disc or ISM morphology, consistent with our findings of MW-like galaxies. However, they showed that CRs significantly influence the phase structure of outflows and the distribution of metals in the CGM. In particular, \citet{Thomas+2024} show that CRs dominate the pressure in the inner CGM and drive turbulent, multiphase winds composed of cold clumps embedded in diffuse, warm gas. \citet{Sike+2025} confirmed these trends and used an idealised tallbox simulation -- allowing for high spatial resolution -- to examine how additional CR transport physics, including nonlinear Landau damping and ion-neutral damping (both also present in \citealt{Thomas+2024}), shape wind properties. Their most complete model, which includes all mentioned damping processes, produces moderate mass loading and substantial CR energy loading, showing that CRs still provide effective feedback despite reduced coupling to the cold ISM. In this model, CRs accelerate warm gas and levitate cool gas in the wind, while having little impact on the coldest and hottest phases. These changes impact key observables, as \citet{Chiu+2024} demonstrated through comparisons between simulations and radio observations.

\citet{Armillotta+2024} used two-moment CR transport in idealised simulations to study how CRs interact with outflowing gas above the galactic disc. They showed that CRs efficiently accelerate warm and warm-hot gas, but have little dynamical effect on fast, hot outflows, where the Alfvén speed is low. Their results emphasize that the thermal state of the gas plays a crucial role in determining how CRs couple to and drive outflows.

These comparisons show that CRs can strongly influence the structure and energetics of outflowing gas, especially in warm and multiphase phases. In MW-like galaxies, our results agree with two-moment studies in finding that CRs have little impact on the star-forming disc and produce low mass loading. This suggests that our one-moment anisotropic diffusion model captures these specific trends. A detailed investigation of how different gas phases are accelerated or affected in our simulations is beyond the scope of this work.

In addition, our cosmological simulations, run with the Auriga model including CR physics, complement the picture by demonstrating that CRs affect the thermodynamic structure of the CGM and extend the results to lower-mass galaxies, where they help to sustain winds as stellar feedback alone is less effective and CR pressure gradients remain steep. As full cosmological simulations with two-moment CR transport remain computationally very demanding, using anisotropic diffusion as an effective model remains a valuable path for studying the role of CRs in a self-consistent large-scale cosmological environment. Future work that combines a cosmological context with more detailed CR transport models will be essential to fully capture the interaction between CRs, multiphase gas, and large-scale outflows.

In addition to two-moment models, some studies have advanced CR modeling by evolving the full CR proton spectrum across multiple momentum bins -- rather than adopting a grey model -- thereby capturing energy-dependent transport and losses. This approach has been used in both isolated and cosmological setups \citep[e.g.,][]{Hopkins+2021a, Ogrodnik+2021, Girichidis+2022, Girichidis+2023, Girichidis+2024}. By resolving the CR energy distribution, these models can self-consistently capture spectral steepening from hadronic and Coulomb cooling, spectral hardening due to transport, and shifts in the effective transport regime across different gas phases. Such energy-dependent effects have been shown to alter the efficiency of CR-driven winds, particularly in dwarf galaxies, and modify the thermal structure and CR pressure support in the CGM of more massive systems \citep{Girichidis+2024}. While these models remain computationally expensive, they provide valuable benchmarks for testing the limitations of simplified CR prescriptions. Comparing results from such spectral models with more approximate treatments can help clarify which physical processes are essential to capture in different regimes of galaxy evolution.

Since CR transport parameters remain uncertain, observational constraints -- such as radio synchrotron and $\gamma$-ray emission -- remain essential for refining CR models in galaxy formation simulations. Comparisons between simulations and observed $\gamma$-ray emission, which directly traces hadronic interactions between CRs and dense gas, have already provided important insights into CR energy densities and confinement across different environments \citep[e.g.,][]{Pfrommer+2017, Chan+2019, Werhahn2021b, Werhahn2023}. However, further observational constraints, particularly on CGM temperature and spatially resolved $\gamma$-ray emission, are still needed to distinguish between competing CR transport and feedback prescriptions in cosmological settings.

\section{Summary and Outlook}
\label{sec:summary}

In this paper, we investigated the role of CRs in galaxy evolution using the SURGE simulations, a suite of cosmological zoom-in simulations based on the Auriga galaxy formation model. Our CR model includes anisotropic diffusion along magnetic field lines, Alfvén cooling that emulates CR streaming via wave damping, as well as Coulomb and hadronic cooling, and instantaneous CR energy injection from newly formed star particles based on a fixed supernova rate and acceleration efficiency. While the thermal energy from supernovae contributes to the effective ISM model, CR energy is injected as a separate channel and evolves according to the CR transport equations.

We explored how CRs influence galaxy morphology, star formation, outflows, and the surrounding CGM across a range of halo masses from dwarf scales to galaxy groups. We also examined the impact of variations in CR transport physics, including different diffusion coefficients and the presence/absence of Alfvén cooling, to assess the robustness of the results to different CR transport parameters. 

Our results highlight the mass-dependent impact of CRs, with lower-mass galaxies being more strongly affected than MW-mass and more massive halos. The results also emphasise the importance of detailed CR transport modeling in galaxy formation simulations. 

Our key findings are:
\begin{itemize}
    \item \textit{CRs affect both disc sizes and star formation rates in galaxies (see Figures~\ref{fig:stellarprojection}, \ref{fig:gasprojection}, and \ref{fig:overview}):} CRs suppress star formation in galaxies with halo masses below $10^{12}\,\Msolar$, reducing stellar mass growth. While CRs also decrease the size of the gas disc and stellar half-light radius for most galaxies, the stellar half-mass radius remains largely unchanged. This is because it reflects the overall stellar distribution, which is dominated by older stars and less sensitive to changes in recent star formation.
    \item \textit{CRs regulate the CGM in lower-mass galaxies (see Figures~\ref{fig:tempprojection} and \ref{fig:profComp}):} CRs modify the temperature, density, metallicity, and magnetic field strength of the CGM in galaxies with halo masses below $10^{12}\,\Msolar$. CR-driven galactic outflows transport CRs, metals, and magnetic fields into the CGM, leading to higher metallicities and lower CGM temperatures in these lower-mass galaxies.
    \item \textit{CRs shape the star formation history of lower-mass galaxies (see Figure~\ref{fig:SFH}):} In galaxies with halo masses below $10^{12}\,\Msolar$, CRs suppress star formation, with the strongest effects in the lowest-mass galaxies, where star formation rates are reduced by up to 50\%. For halo masses above $10^{11.5}\,\Msolar$, the impact of CRs on star formation diminishes, becoming negligible in the most massive galaxies. In the intermediate-mass regime ($10^{11} - 10^{11.5}\,\Msolar$), star formation varies stochastically across simulations, making it harder to isolate the CR contribution. The effect of CRs may still be present but is small compared to this inherent variability.
    \item \textit{The impact of CRs on outflows depends on halo mass (see Figures~\ref{fig:velslice} and \ref{fig:OutflowMassLoading}):} In lower-mass galaxies, CRs sustain winds over larger distances, increasing mass outflow rates and velocities at the virial radius, despite initially weaker winds near the galaxy compared to simulations without CRs. In contrast, we find no noticeable effect of CRs on outflow properties in MW-mass and more massive galaxies, likely because strong winds are already driven by stellar and AGN feedback in these systems.
    \item \textit{Varying CR transport properties affects galaxy evolution, particularly in lower-mass systems (see Figures~\ref{fig:Xcr}, \ref{fig:VariationsDisk}, \ref{fig:VariationsOutflow} and \ref{fig:VariationsSFH}):} Increasing the CR diffusion coefficient leads to higher outflow velocities but lower mass-loading factors in low-mass galaxies, because CR energy is transported more efficiently away from star-forming regions. Additionally, turning off Alfvén cooling enhances mass-loading factors in the extended CGM for both intermediate- and low-mass galaxies, underscoring the role of CR energy losses in shaping winds. Because of their strong sensitivity to variations in CR transport parameters, low-mass galaxies provide valuable environments for testing and refining CR transport models.
\end{itemize}

In future work, we plan to investigate the role of CRs across cosmic time to understand how their effects evolve. Our results motivate the development of improved CR transport prescriptions for cosmological simulations with lower resolution than the isolated galaxy simulations that can afford to include a two-moment CR transport model. In addition, extending our simulation suite to include more galaxies across a broader range of halo masses and accretion histories, and/or simulations in larger volumes, will help disentangle the role of CRs from cosmic variance and differences in assembly histories. Finally, observational comparisons -- particularly using $\gamma$-ray and radio synchrotron emission -- will be essential not only for constraining CR transport parameters and refining CR feedback models, but also for improving galaxy formation models more broadly.

\section*{Acknowledgements}

RB is supported by the UZH Postdoc Grant, grant no. FK-23116 and the SNSF through the Ambizione Grant PZ00P2\_223532.
FvdV is supported by a Royal Society University Research Fellowship (URF\textbackslash R1\textbackslash191703 and URF\textbackslash R\textbackslash241005). CP acknowledges support by the European Research Council under ERC-AdG grant PICOGAL-101019746.

\section*{Data Availability}
The data underlying this article will be shared on reasonable request to the corresponding author.



\bibliographystyle{mnras}
\bibliography{surgecr} 



\appendix



\bsp	
\label{lastpage}
\end{document}